\documentclass{article}

\usepackage{arxiv}
\usepackage[utf8]{inputenc} % allow utf-8 input
\usepackage[T1]{fontenc}    % use 8-bit T1 fonts
\usepackage{hyperref}       % hyperlinks
\usepackage{url}            % simple URL typesetting
\usepackage{booktabs}       % professional-quality tables
\usepackage{amsfonts}       % blackboard math symbols
\usepackage{nicefrac}       % compact symbols for 1/2, etc.
\usepackage{microtype}      % microtypography
\usepackage{graphicx}
\usepackage{doi}
\usepackage{amsmath}
\usepackage{physics}

\providecommand{\keywords}[1]{\textbf{\textit{Index terms---}} #1}

\usepackage{xcolor}
\usepackage{soul}
\usepackage{graphicx}

\newcommand{\authors}[2]{%
	\begin{center}
		{{\large #1} \\ \vspace*{0.3cm}{#2}}
\underline{}\end{center}}

\title{Coherent all X-ray four wave mixing at core shell resonances}

\date{}

\renewcommand{\headeright}{}
\renewcommand{\undertitle}{}
\begin{document}
\maketitle

\vspace*{-2cm}

\authors{A.S. Morillo-Candas$^{1,*}$, S.M. Augustin$^1$, E. Prat$^1$, A. Sarracini$^1$, J. Knurr$^1$, S. Zerdane$^1$, Z. Sun$^1$, N. Yang$^{1,2}$, M. Rebholz$^3$, H. Zhang$^{1,2}$, Y. Deng$^1$, X. Xie$^1$, A. Cannizzo$^4$, A. Al-Haddad$^1$, K.A. Schnorr$^1$, C. Ott$^3$, T. Feurer$^4$, C. Bostedt$^{1,2}$, T. Pfeifer$^3$, G. Knopp$^{1,*}$ }
{$^{1}$ \textit{Paul Scherrer Institute, Forschungsstrasse 111, Villigen, CH-5232, Switzerland} \\
$^{2}$\textit{École Polytéchnique Fedérale de Lausanne, Lausanne, CH-1015, Switzerland} \\
$^{3}$\textit{Max-Planck-Institut für Kernphysik, Saupfercheckweg 1, Heidelberg, 69117, Germany} \\
$^{4}$\textit{Institute of Applied Physics, University of Bern, Sidlerstrasse 5, Bern, CH-3012, Switzerland} \\
\vspace*{0.2cm}
$^{(*)}$ \textit{ana.morillo@psi.ch, gregor.knopp@psi.ch}}

\vspace*{1cm}

\begin{abstract}
Nonlinear wave mixing in the X-ray range can provide %promises 
valuable insights into the structural and electron dynamics of atomic and molecular systems on ultrafast time scales, with state- and site-selectivity and atomic resolution. This promising experimental toolbox was so far limited by requiring at least one near-visible laser, thus preventing core-shell two-dimensional X-ray spectroscopy. In this work, we demonstrate the generation of background-free all-X-ray four-wave mixing (XFWM) signals from a dilute gaseous sample (Ne). The measured and simulated two-dimensional spectral maps ($\omega_{\text{in}},\omega_{\text{out}}$) show multiple contributions involving the coherent response from core electrons. Notably, two-color resonant XFWM signals, essential for generalized multi-color schemes that allow to locally probe the electronic excitation of matter, are observed in neutral Ne. Moreover, stimulated Ne$^+$ emission in each of the propagating X-ray pulses leads to an increase of the temporal coherence in a narrow-bandwidth, which results in the coherent mixing of three X-ray lasers. Preliminary X-ray excitation experiments making use of multi-color time-delayed X-ray pulses demonstrate temporal resolution capability and show a time dependency consistent with a signal dominated by resonant XFWM processes. 
This first all-X-ray four-wave-mixing approach \cite{morillo2023allXray} represents a major breakthrough towards multidimensional X-ray correlation spectroscopy and the general application of nonlinear all-X-ray wave-mixing.
\end{abstract}

% keywords can be removed

\section{Introduction} \label{Sec:Introduction}

\par Ultra-fast inner-shell electron dynamics and electron correlations are fundamental to characterize atomic and molecular states, their structure and dynamics, which are at the heart of chemical reactions and biochemical processes \cite{ott2014reconstruction, oneal2020electronic}. The concerted motion of electrons and atoms defines the electronic and magnetic properties of matter and is fundamental for multiple phenomena ranging from phase transitions \cite{ossiander2017attosecond} to manipulating quantum coherence in atomic systems, relevant for quantum information and energy transfer \cite{feynman1982international, saffman2010quantum}. To understand and control molecular 
quantum systems, information on the local excitation, site-dependent energy- and charge transfer involving coherent pathways is essential. However, relaxation processes, where dissipation and coherence properties become intermixed, gives rise to complex dynamics, often inaccessible by linear light-matter interaction techniques. 

\par Nonlinear wave mixing at the atomic scale can provide multi-dimensional information about structural and electronic dynamics \cite{mukamel2005multiple,mukamel2013multidimensional}. State selectivity and quantum control 
can be achieved by sequences of light-matter interactions at which particular sub-processes are enhanced by matching the photon frequencies to resonant transitions between specific atomic and/or molecular states \cite{avisar2011complete,ott2014reconstruction,zhang2016nonlinear}. The extension of wave mixing techniques to the X-ray range would inherently enable atomic scale (nm) spatial resolution at ultrafast time scales (femtosecond to attosecond), whereas X-ray excitation in resonance with core and/or inner shell electron transitions, allows element specificity and site-selective excitation. In particular, X-ray four-wave mixing (XFWM) can provide key insights into the role of coherence, correlation and interference \cite{fano1961quantum,meier1995signatures}. However, these powerful experimental approaches have been so far limited to the mixing of X-ray beams with near-visible/optical lasers, \cite{glover2012xray}, not being able to fully exploit the multiple advantages of X-ray spectroscopy. 

\par The application of nonlinear multi-photon techniques in the X-ray range is indeed challenging. The interaction of X-ray photons with electrons from core-shell orbitals typically yields short-lived core-excited species. Short radiative lifetimes and small transition cross-sections are disadvantageous against competing processes, such as single- or multi-photon ionization and, particularly, cascading relaxation via non-radiative Auger-Meitner (A-M) decay. In addition, these processes generate multiple product species that competitively interact with the incoming photons and often imply a severe material damage. The study of nonlinear phenomena in the X-ray regime thus had to await for the short pulse durations, closer to the excited states lifetimes, and high intensities produced at X-ray Free-Electron Lasers (XFELs). Following a series of experiments in the extreme ultraviolet (EUV) \cite{bencivenga2015four, maznev2018generation, bohinc2019nonlinear, bencivenga2023extreme}, the first FWM signals involving multiple X-ray photons were recently measured in the form of X-ray/optical mixed transient gratings (XOTG). In these experiments, the X-ray induced periodic modulation of the refractive index was probed optically, measuring the partially incoherent energy transfer to phonons, heat and pressure waves or changes in the chemical environment \cite{maznev2018generation, rouxel2021hard, morillo2023liquid}.

\par FWM approaches that involve solely X-ray photons are particularly interesting. Coherent four-wave mixing addressing core resonances is the key to access the intra-system dynamics of matter. Used in a two-dimensional spectroscopy configuration, a sequence of multiple X-ray FEL pulses can create correlated electron wavepackets, revealing the coupling between distinct quantum states. It would also allow to understand the role of coherent electron motion in a wide range of charge and energy transfer phenomena between atoms within a system, making it possible to investigate the nonlocal nature of electronic excitations \cite{tanaka2002coherent,oneal2020electronic}. In addition, X-ray gratings can probe transient periodicities in the nanometer range, providing information on the collective relaxation and transport phenomena at nanoscale \cite{bencivenga2019nanoscale,bencivenga2023extreme}. Multiple intra- and inter-atomic relaxation pathways in out-of-equilibrium atomic or molecular systems can thus be addressed by all-X-ray FWM. As such, it has been envisioned and theoretically described some time ago \cite{tanaka2002xray,mukamel2005multiple,mukamel2013multidimensional} but remains unaccomplished, presenting a long-standing experimental goal. In this work, we demonstrate all-X-ray four-wave mixing (XFWM) in an atomic system, neon \cite{rohringer2012atomic, weninger2013stimulatedElectronic,eichmann2020photon,morillo2023allXray}, as a first step towards the general application of coherent all X-ray multi-wave spectroscopies.  

\section{All X-ray FWM signal}

\begin{figure}[t]
	\centering
    % \fbox{
   	\includegraphics[width=0.95\textwidth,trim=0cm 11cm 0cm 11.8cm,clip]{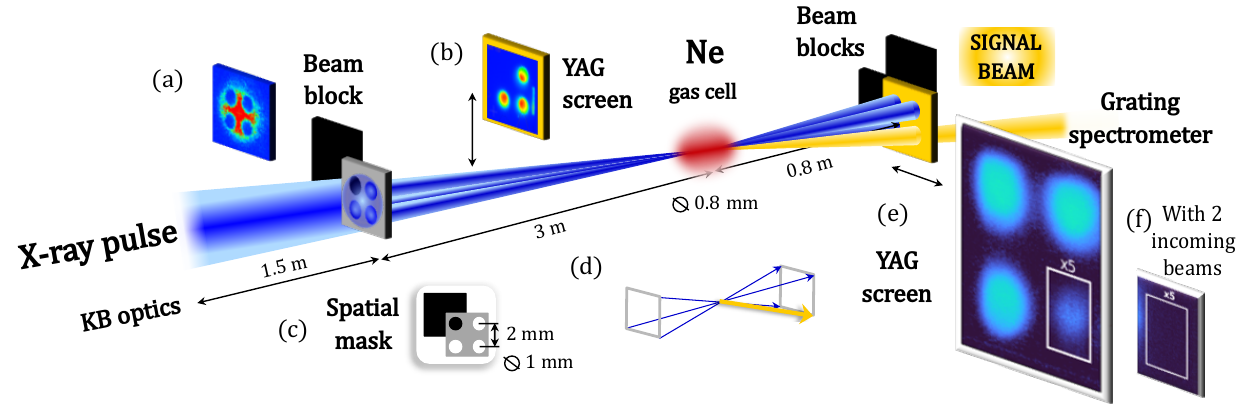}
    % }
	\caption{\textbf{Scheme of the experimental setup and spatial characterization of the XFWM signal}: (a) Camera image of an optical laser with similar beam size as the incoming focused FEL beam on the spatial mask to represent the imprint of the FEL on the mask. A beam block closes one of the four holes directly after the mask; (b) Camera image of the fluorescence from a Ce-YAG screen located after the mask showing the three cut-out X-FEL beams; (c) Scheme and dimensions of the spatial mask; (d) Scheme of the beams distribution before and after the focus in the `folded Box' geometry. Beam propagation around the focus and intensity distribution at the focus are given in the supplementary information \ref{sec:SImaskTransmission}; (e) Camera image of the fluorescence from a Ce-YAG screen placed after the gas cell, showing the spatial distribution of the three incoming beams and the signal beam, amplified by a factor of 5. The displayed signal is measured for a photon energy of $\sim\,$867 eV, corresponding to the maximum of the signal (for other photon energies see the SI \ref{sec:SIyag}). With only two incoming beams no signal is detected (f). When the screen is removed, the signal beam reaches a grating spectrometer.}
	\label{fig:ExpSetup}
\end{figure}

The all-XFWM experiment was performed using the so-called `folded Box' or `BoxCARS' configuration, in which three beams with wavevectors $\Vec{k}_1$, $\Vec{k}_2$, $\Vec{k}_3$ are focused from positions corresponding to three corners of a square \cite{eckbreth1978boxcars,prior1980three}. Energy $\sum_{n}{\hbar \omega_{n}} \approx 0 $ and momentum conservation $\sum_{n}{\hbar \vec{k}_{n}} \approx 0 $ define the frequency and direction of the emitted photons, separated from the incoming beams and therefore allowing a background-free signal detection. To implement the `Box' configuration in the X-ray range, we use a spatial mask that splits a single focusing beam into three `cut out' sections that propagate towards the focus (Fig.$\,$\ref{fig:ExpSetup}), where a gas cell filled with a few hundred mbar of neon is positioned. In this robust configuration, and without any further modification, temporal and spatial overlap at the crossing point are given and the phase-matching condition ($\Vec{k}_{s}\,=\,\Vec{k}_1\,-\,\Vec{k}_2 +\,\Vec{k}_3$) is automatically fulfilled \cite{belabas2005four}. The experiment was performed at the Maloja end-station of the Swiss Free Electron Laser (SwissFEL), which provides soft X-ray pulses produced by self-amplification of spontaneous emission (SASE) with a standard pulse duration of 25 fs RMS and a pulse energy of $\sim\,$2 mJ for photon energies around 870 eV \cite{prat2023xray}. The transmission of the mask is estimated to be $\sim\,$14\%, which corresponds to a total on-target photon flux of $\sim\,$2$\times$10$^{12}$ photons per pulse.

\par XFWM photon scattering into the phase-matched direction was evidenced by imaging the fluorescence from a Ce-YAG screen inserted into the beam path after the sample. The YAG images show a well defined round-shaped signal beam profile, a `laser-like' signal, at the expected position, isolated from the three incoming beams and with a similar divergence (Fig.$\,$\ref{fig:ExpSetup}$\,$(e)). The experiment thus highlights one of the major advantages of the non-collinear `Box' geometry. Signal scattering into a background-free direction allows to distinguish the XFWM signals from the incoming beams and other co-propagating non-phase matched nonlinear contributions, facilitating a detailed spectral analysis. 

\par No signal emission into the phase-matched direction was observed when blocking any of the incoming beams before the interaction region (see Fig.$\,$\ref{fig:ExpSetup}$\,$(f) and the supplementary information, \ref{sec:SIyag} and \ref{sec:SIspectra}). Three beams were needed to generate the measured signals. This verification along with the observed signal direction and divergence support the successful generation of all X-ray FWM signals. The efficiency of the nonlinear process, defined as the ratio of the maximum number of signal photons and the incoming beams is in the order of 0.15\%. The strength of the signal is therefore particularly remarkable, visible on a single shot basis on a fluorescence screen.

\section{Signal generation and model}  

\par
\begin{figure}[t]
	\centering
    % \fbox{
 	\includegraphics[width=1\textwidth, trim=1cm 10cm 0.8cm 9cm,clip]{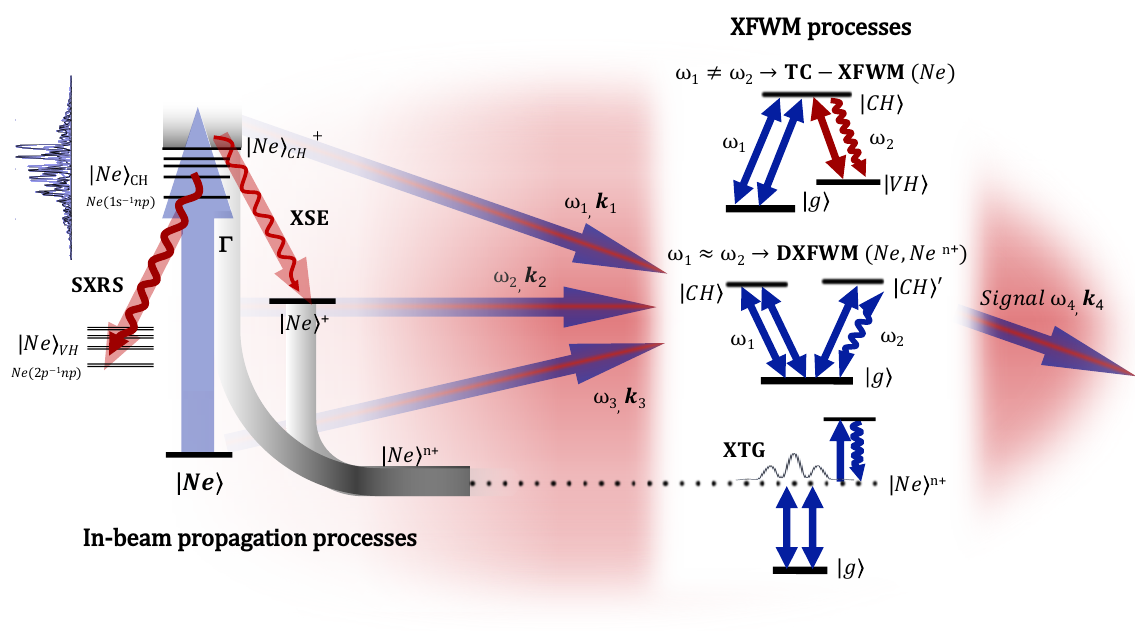}
    % } 
	\caption{ \textbf{Processes involved in the XFWM signal formation:} Schematic representation of different processes that contribute to the measured FWM signals, including in beam build-up propagation of frequencies (XSE and XSRS), to the left, that are scattered into the phase-matched direction \textit{via} their contribution to resonant XFWM (DXFWM or SRXFWM) or non-resonant (XTG) processes, to the right (CH: core-hole; VH: valence-hole). These processes are included in the model developed to interpret the measured spectra.}
	\label{fig:Discussion}
\end{figure}

 The total XFWM signal is proportional to the square of the induced third order polarization, $\abs{P^{(3)}(\omega,\vec{r})}^2$, in an homodyne detection scheme. Resonant and non-resonant processes, summarized in Fig.$\,$\ref{fig:Discussion}, can contribute to the XFWM signals. Different resonant XFWM excitation schemes are possible due to the large bandwidth (7.8 eV) and multi-mode structure of the SASE FEL pulses \cite{williams1997two}. Two-color, inelastic, resonant contributions (TC-XFWM) involve two interactions with similar energy that resonantly populate the $np$ core-excited states, while the 3$^{rd}$ interaction is resonant with a core-to-valence transition \cite{weninger2013stimulatedElectronic,oneal2020electronic}. The signal photon is thus red-shifted (Raman-type scheme) and the process results in a non-parametric transfer of the electronic population to a valence-excited neutral state of neutral neon. A second possible excitation pathway couples the common ground state with either the same or two distinct, but energetically close, core-excited states. All species with transitions in the range of the explored photon energies can produce this type of resonant \textit{quasi} `degenerated' XFWM signals, denoted DXFWM in the following.

\par Non-resonant contributions are dominated by ion gratings. The high ionization probability in the soft X-ray range along with the spatial intensity distribution given by the interference of the three focusing X-ray beams creates a spatially modulated ion and electron density all along the focal region (see the SI \ref{sec:SImaskTransmission}). The X-ray induced transient volume grating (XTG) rapidly formed within the pulse duration can non-resonantly scatter photons from the incoming beams. Transient gratings formed in time scales longer than the pulse duration, \textit{e.g.}, induced by heat, pressure, etc., do not contribute relevantly to our signals, produced within the pulse. 

\par The strength and resulting spectra of XFWM processes depend on the available spectral photon density in the absorption and emission bandwidths. Propagation effects thus play a significant role in the nonlinear interaction of the incoming X-ray beams with the sample, \textit{via} the coherent build-up of frequencies in the interaction volume \cite{Stohr2015,rohringer2012atomic}. Stimulated X-ray emission (XSE) and stimulated X-ray Raman scattering (SXRS) modify the spectral density and coherence of each of the beams along its propagation pathway ($\Vec{k}_i$) \cite{rohringer2012atomic,weninger2013stimulatedElectronic}, and therefore partially define the frequencies observed in the XFWM spectra detected in the phase-matched direction ($\Vec{k}_1 -\Vec{k}_2 + \Vec{k}_3$). Note that SXRS and inelastic TC-XFWM are both third order non-linear processes based on comparable excitation energy levels schemes. However, SXRS is proportional to the imaginary part of $P^{(3)}$ and is generated and amplified in beam direction. In contrast, TC-XFWM is proportional to $\abs{P^{(3)}}^2$ and the signal is generated where the beams intersect. Additionally, SXRS implies strictly the interaction of photons with two different energies, whereas TC-XFWM can be generalized to a more versatile multi-colors case, with different energies for all four interactions involved. 

\par Resonant and non-resonant contributions are considered in a model developed to support the interpretation of the data. The model describes the different processes in the frequency domain \textit{via} expressions derived from perturbation theory \cite{kramers1925streuung,williams1997two,roslyak2009nonlinear}, but considering the time-dependent population and spectral changes within the SASE pulse duration. Propagation effects are calculated by solving the wave equation in its linearized form accounting for linear (related to X-ray absorption and XSE) and nonlinear polarization terms \cite{rohringer2009atomic, weninger2013stimulatedResonant} (more details on the model can be found in the supplementary information \ref{sec:SIsignalGeneration}). 

\begin{figure}
	\centering
    % \fbox{
 	\includegraphics[width=0.8\textwidth,trim=1.25cm 4.5cm 1.75cm 1.5cm,clip]{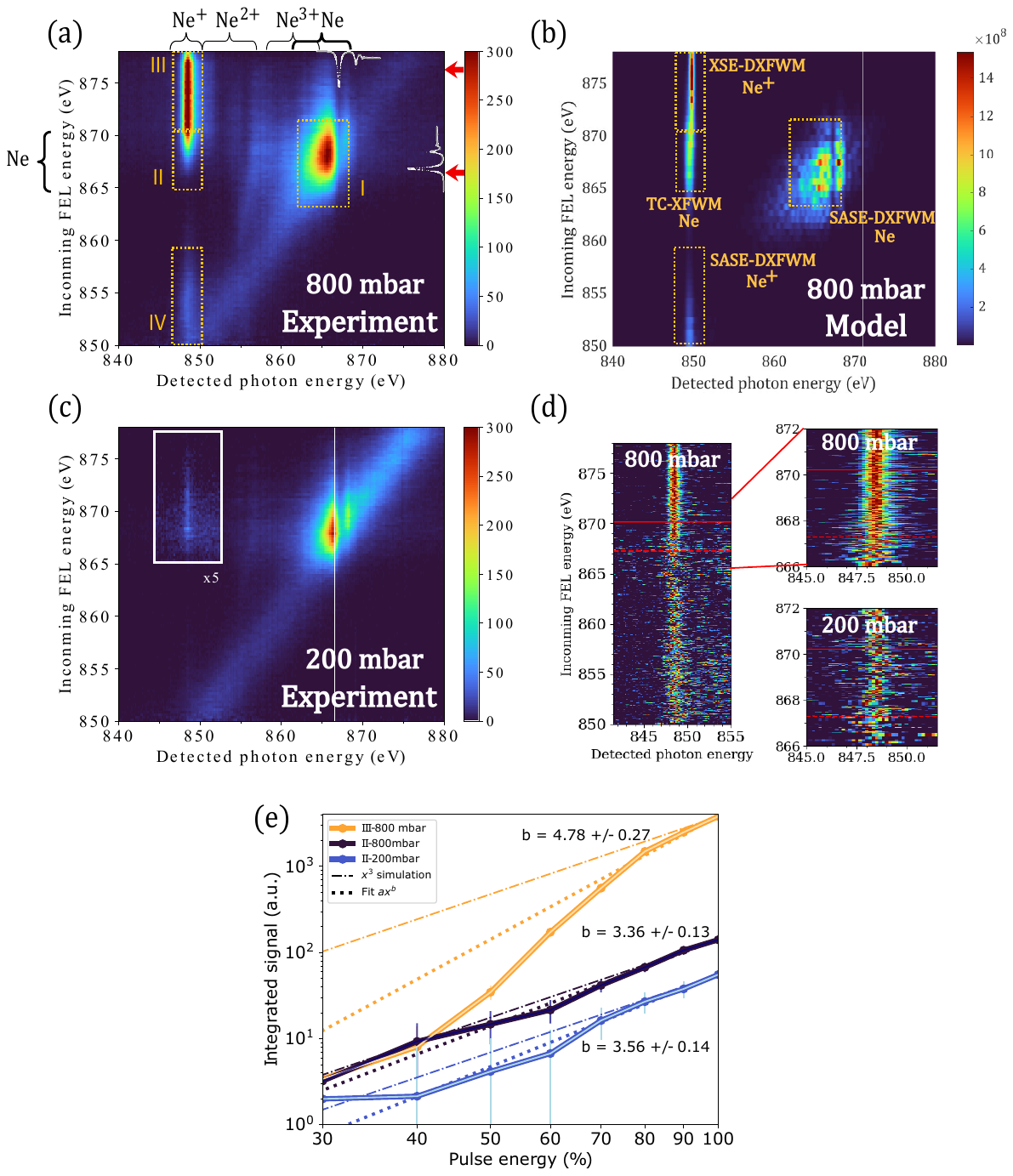}
    % }
	\caption{\textbf{Spectral analysis and simulation of the all X-ray four-wave mixing signal:} (a) Average spectra ($\sim\,$600 shots/row) measured at 800 mbar as a function of the incoming and emitted photon energies. Four Regions of Interest (RoIs) of the map are highlighted with dashed boxes and numbered from I to IV. Absorption spectra are plotted as overlapping white curves, as reference; (b) Simulated 2D spectral map (shot-to-shot) of the XFWM signal calculated in the same conditions as in panel (a), indicating the main processes and species involved in the signal generation for the 4 RoIs;   (c) Average spectra measured at 200 mbar. The well isolated TC-XFWM signal in Ne is amplified by a factor of 5; 
    (d) Single shot spectra normalized to the maximum of the signal around 848.5 eV for the 800 mbar Ne case. The two zoom plots show the spectra around $\omega_{in}\approx$867-870 eV for both 800 and 200 mbar, for the most intense peaks (10 times the noise level);
    (e) Variation of the integrated signals detected around $\omega_{out}$=848.5 eV as a function of the pulse energy and measured for 800 and 200 mbar at two incoming photon energies (866.4 and 876.6 eV), highlighted with red arrows in panel (a). These integrated signals are representative of the TC-XFWM signal in Neon and the XSE-DXFWM signal in Ne$^{+}$ (RoIs II and III in panel (a) respectively). Fits according to a power law $ a \cdot x^b$ and simulations of a power 3 curve are included as reference. The error is calculated as the standard deviation within all the recorded shots. }
	\label{fig:Spectra&Fits}
\end{figure}

\section{XFWM spectra} \label{sec:spectra}

\par The 2D XFWM maps as a function of incoming and emitted photon energies ($\omega_{in}$-$\omega_{out}$) show notably rich spectra with a manifold of spectral features given by the multiplicity of processes and species involved. Two representative cases are presented in Fig.$\,$\ref{fig:Spectra&Fits} measured at 800 and 200 \noindent mbar Ne pressure and 25 fs RMS pulse duration. The main spectral features are highlighted as Regions of Interest (ROIs) in the figure; all of them show clear nonlinear dependencies when scanning the FEL intensity (see Fig.$\,$\ref{fig:Spectra&Fits}$\,$(e)). 

\par The involved transitions can be readily identified from the available literature. The explored FEL photon energies between 845-880 eV cover the pre-edge resonances of Ne, transitions of 1s core electrons to the empty np orbitals (n = 3, 4, 5, 6, between 867.3 and 869.8 eV), and the direct core-ionization threshold (Ne K absorption edge, at 870.2 eV). The lifetime of core-excited states is limited by the A-M decay to 2-3 fs \cite{weninger2013stimulatedResonant, muller2017photoionization} and leads to the production of valence ions, which are also generated by direct ionization from the 2s and 2p orbitals for all photon energies. In the explored frequency range, transitions from three ions contribute to the measured signals: core-excited Ne$^{+}$ emits photons centered around $\omega_{out}$=848.5 eV, whereas the photo-emission from single core-hole Ne$^{2+}$ and Ne$^{3+}$ falls into the ranges of 850-856 eV and 859-865 eV respectively \cite{agren1978multiplet, obaid2018lcls}.  

\par The simulated 2D XFWM spectral map for the 800 mbar Ne case is shown in Fig.$\,$\ref{fig:Spectra&Fits}$\,$(b). The model captures all the relevant spectral features for the considered species, Ne and Ne$^{+}$, qualitatively reproducing the measured XFWM signal. The model thus provides information on the dominant signal generation mechanism, indicated next to the corresponding RoI along with involved species.

\section{Interpretation of the XFWM spectra} \label{sec:spectralChar}

\par Two contributions in the 2D-map relate to neutral neon. For incoming photon energies around the Ne pre-edge resonances (867-869 eV), we observe a strong broad signal for $\omega_{out}$ around the 1s-to-np transitions (RoI I in Fig.$\,$\ref{fig:Spectra&Fits}$\,$(a)) and a Raman-shifted signal around $\omega_{out}\sim\,$848.5 eV (RoI II). The latter is particularly evident at lower pressures where it is clearly visible and well isolated (see for 200 mbar Ne in Fig.$\,$\ref{fig:Spectra&Fits}$\,$(c)).  The first of these contributions is generated \textit{via} resonant DXFWM process of the incoming SASE beams (SASE-DXFWM), whereas the phase-matched Raman-shifted signals are generated by two-color resonant XFWM (TC-XFWM) involving electronic Raman transitions. These signals, for 200 and 800 mbar, show the expected third order intensity dependence ($\propto I^3$) of a FWM process (Fig.$\,$\ref{fig:Spectra&Fits}$\,$(e)). 

\par For both SASE-DXFWM and TC-XFWM, the stochastic spectral distribution of photons in the incoming SASE pulse defines the observed spectra on a shot-to-shot basis. The detuning of the multiple spectral spikes of the SASE pulse from the resonances leads to spectral fluctuations, visible in the single shot spectra shown in Fig.$\,$\ref{fig:Spectra&Fits}$\,$(d). These fluctuations are a signature of the interaction of broadband multi-mode X-ray radiation with resonances and allow to distinguish this type of signals from those originated in stimulated emission processes \cite{weninger2013stimulatedElectronic, weninger2013stimulatedResonant}. 

\par All signals stemming from the ions are a consequence of resonant DXFWM processes. The involved photons originate either from the incoming SASE pulses (SASE-DXFWM) or from stimulated frequencies that build up along beam propagation (XSE-DXFWM). Both cases show different spectral properties, particularly visible in the Ne$^{+}$ signals at 848.5 eV, highlighted with RoIs III and VI in Fig.$\,$\ref{fig:Spectra&Fits}$\,$(a). For incoming FEL energies below the K-edge, the mechanism is similar as for neutral neon and therefore the SASE-DXFWM signal exhibits the expected spectral fluctuations correlated with the incoming SASE pulses (Fig.$\,$\ref{fig:Spectra&Fits}$\,$(d)). This signal is maximum when the FEL photon energy is centered at the transition energies of valence ions ($\omega_{in}$=$\omega_{out}$) and vanishes when the FEL moves away from the resonances (see also Fig.$\,$\ref{Fig:SIlowEnergy}). 

\par For photon energies above the ionization edge, each of the incoming beams generates amplified stimulated emission from core-excited Ne$^+$, produced by the direct core-ionization of Ne \cite{rohringer2012atomic, weninger2013stimulatedElectronic}. As a consequence, each of the incoming fields increases its temporal coherence and spectral density in a narrow bandwidth range (0.27 eV) around 848.5 eV. The interaction of the three converging `laser-like' beams leads to a strong coherent all-X-ray FWM signal at the same energy (XSE-DXFWM) characterized by a well defined, narrow spectrum at a fixed frequency and with a power dependency $\sim\,I^{>4}$, given by the XSE amplification along the beam propagation (see Fig.$\,$\ref{fig:Spectra&Fits}$\,$(e)). In contrast to SASE-DXFWM, where the signal frequencies reflect the spectra of the incoming beams, for high photon energies in-beam propagation XSE leads to X-ray lasing \cite{rohringer2012atomic}. The intense XSE-DXFWM signal is thus the result of coherent four-wave mixing with three narrow bandwidth X-ray lasers.

\section{Time-resolved X-ray four-wave mixing and outlook} \label{sec:spectralChar}

\par All X-ray FWM has been demonstrated and proven as a powerful
spectroscopic tool. Resonant XFWM contributions dominate the 2D-spectra. Both neutral neon and its ions contribute to generating coherent signals when the incoming radiation matches their specific core-to-valence transitions. The observed two-color XFWM signals involving electronic Raman transitions are particularly important to establish generalized multi-color coherent Raman FWM schemes in the X-ray regime. In the future, tuning the difference of two X-ray frequencies across the valence excitations will open the door for the direct probing of the entire manifold of electronic states and coherences in atomic or molecular systems \cite{tanaka2002coherent}. 

\par
\begin{figure}[t]
	\centering
    % \fbox{
    \includegraphics[width=0.85\textwidth,trim=1cm 8cm 1cm 8cm,clip]{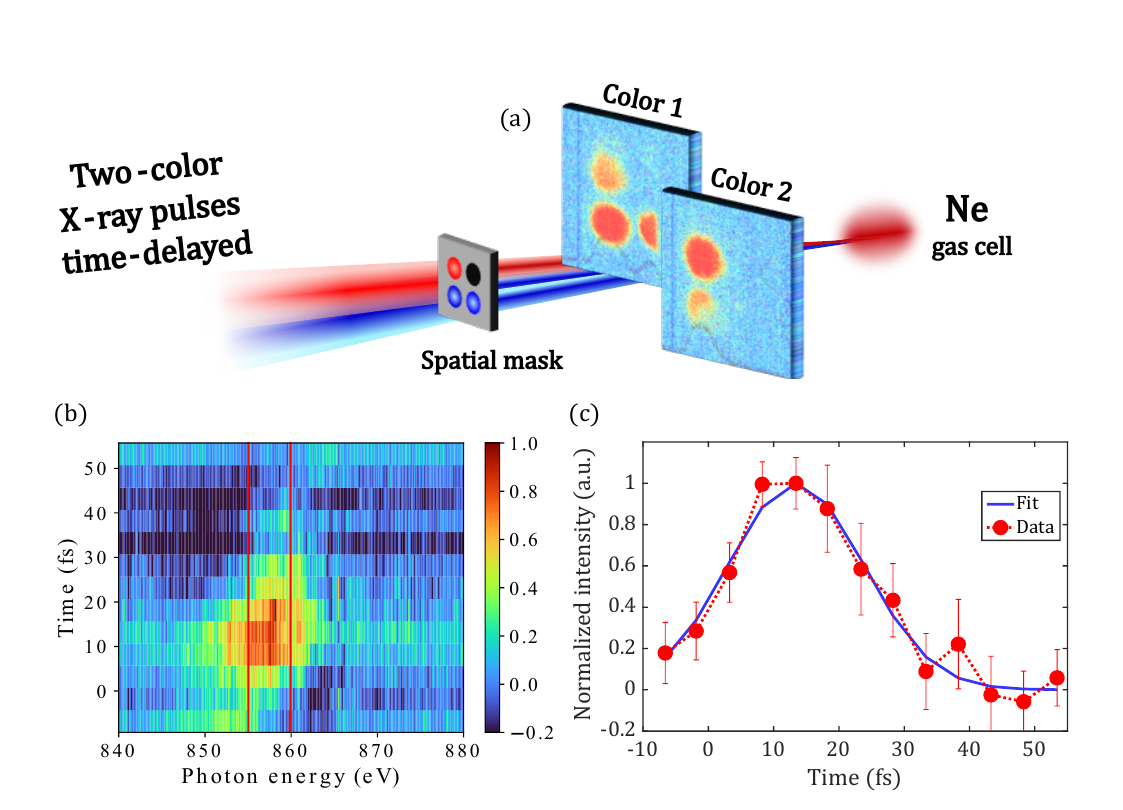}
    % }
	\caption{\textbf{Two-color X-ray four-wave mixing:} Scheme of the experimental setup for the two-color mode with the camera images recording the fluorescence from a Ce-YAG screen located 0.24 m away from the spatial mask and (a) showing the remaining FEL beams with either color 1 only to the left (resonant pump at $\sim\,$866 eV), or color 2 only to the right (red-shifted probe at $\sim\,$858 eV); (b) Difference spectral plot as a function of the detected photon energy and time delay between pump and probe pulses; (c) Integral of the signal as a function of the time delay and fit. The plotted signal is the sum of the signal within the red lines shown in panel (b). The data is fitted as the convolution of the pump and probe pulses with a single exponential decay function corresponding to the 2.5 fs A-M life time. The fit estimates a pulse duration $\sim$$\,$24 fs FHWM (10 fs RMS), consistent with the shorter gaussian pulses given by the two-color mode. 
    }
	\label{fig:twoColors}
\end{figure}

 XFWM time-resolved measurements with specific time ordering of the interacting pulses will be needed to monitor the electron dynamics induced by the coherent superposition of states. Moreover, different XFWM contributions, often indistinguishable in the frequency domain, can be disentangled via distinct excitation pathways in the time domain. As a first attempt of time-resolved all-XFWM, a set of preliminary time-resolved experiments have been realised making use of the two-color X-ray capabilities of the Athos branch of SwissFEL in combination with the `Box' configuration (Fig.$\,$\ref{fig:twoColors}). These exploratory  measurements confirm the observation of clear time-resolved signals. They also confirm that non-resonant processes given by the elastic scattering from an ion X-ray transient grating (XTG) are of minor importance in our experimental conditions, and not essential to explain any of the main spectral features. XTG signals are expected to show a slow decay as a function of time, in the picosecond to nanosecond time scale, given by the speed of diffusion processes \cite{rouxel2021hard,morillo2023liquid}. However, the time-resolved XFWM signal fully vanishes after $\sim\,$30 fs, reflecting the temporal convolution of pump and probe pulses (estimated to be up to 10 fs RMS pulse duration in the two-color mode) and the 2.5 fs excited-state lifetime given by the A-M decay (Fig.$\,$\ref{fig:twoColors}$\,$(c)), as expected from resonant XFWM signals involving short-lived core-excited states. Future experiments with time resolution down to few fs or shorter will allow to resolve the dynamics of (multi-)core-hole states.

\par The strength of the ion contributions to the XFWM signals is particularly relevant in the context of coherent FWM spectroscopy in the X-ray range. In general, stochastic relaxation processes, such as A-M decay, are believed to severely inhibit coherent FWM signal formation, since they typically lead to substantial ion production. However, these ions may play, on the contrary, a favorable role in the generation of otherwise weak nonlinear XFWM signals. Stimulated radiation from the multiple ion species can substantially increase the photon density in a broader spectral range and thus stimulate weak transitions also in neutral species. A strong local oscillator field in homodyne detected FWM spectroscopy can potentially enhance the detection of weak signals, similar to heterodyne techniques. Coherent signals originating in the ions (\textit{e.g.}, XSE-DXFWM) or photons Bragg-scattered by the ion gratings (XTG) could propagate together with weak resonant XFWM contributions into the same phase-matched direction, leading to optical and quantum interference in the third order response. Finally, X-ray lasing processes in-beam propagation can increase the available photon energies and thus facilitate the implementation of complex multi-color experimental schemes.

\par The capability to perform all X-ray FWM measurements in the time domain and the large experimental flexibility provided by varying experimental parameters such as propagation length, pressure, sample composition or the intensity distribution at the focus to have a grating pitch comparable with the short FEL wavelength \cite{rouxel2021hard, bencivenga2019nanoscale}, can be used to enhance or inhibit certain contributions to the FWM signal and to disentangle the ultrafast coherent and incoherent dynamics of the system. Both will make it possible to fully exploit the spectroscopic potential of all-X-ray four-wave mixing in numerous fields, from material science to biochemistry.

\par In summary, we have demonstrated the generation of strong coherent all-X-ray FWM signals using a very robust approach that ensures temporal and spatial overlap of the incoming pulses and phase-matching. The signal spectrum emitted in phase-matched direction is dominated by resonant contributions on neutral Ne and its ions, including inelastic Raman two-color XFWM signals, the keystone for generalized multi-color XFWM schemes. Preliminary results exploring the two color-mode show the capability to detect transient signals and demonstrates the potential to extend the methodology to the time domain as a general multi-colour and time-resolved all-X-ray four-wave mixing spectroscopic tool. The presented experiment therefore represents a crucial step towards the general application of nonlinear all-X-ray wave-mixing experiments and towards 2D X-ray spectroscopy and paves the way to monitor valence electron wavepackets and the investigation of correlated electron dynamics.

\section{Methods}  \label{sec:methods}

\small

\subsection{Set-up and experimental details}

\par The gaseous sample for the experiment, Ne (Linde 5.0, 99.999\%), is delivered at the focus of the X-ray beam using an in-vacuum cylindrical gas cell of 6 mm inner diameter, 8 mm outer diameter, filled with pressures in the range of 200 to 800 mbar (5$\times$10$^{24}$ to 2$\times$10$^{25}$ atoms per m$^{3}$). The spatial mask is a 1 mm thick aluminium plate with 4 identical holes of 1 mm diameter, geometrically arranged in the corners of a 2$\times$2 mm square. The size of the spatial mask was chosen in a way to find a good trade-off between introduced angle between the k-vectors and transmission through the mask. The mask was inserted into the path of the soft-X-ray beam after the focusing Kirkpatrick-Baez (KB) mirrors of the Maloja end-station (Fig.$\,$\ref{fig:ExpSetup}). The transmission of the mask was estimated assuming a propagating Gaussian beam (more details in \ref{sec:SImaskTransmission}). Behind the mask, one of the holes is permanently blocked, while the other holes can be flexibly opened or closed in order to verify that the XFWM signals are generated only when all three incoming beams are present.

\par The directed background-free signal was detected with an in-line Scienta X-ray spectrometer (Scienta XES350, resolution $\Delta$E$\approx$0.4 eV). The incoming X-ray beams are monitored shot-to-shot with a hemispherical electron analyser (Specs, Phoibos 150 EP,  resolution $\Delta$E$\approx$0.3-1.1 eV) located between the mask and the interaction region (reference spectra) that allows to characterize the central photon energy and intensity. The gas used in the electron spectrometer chamber was argon and the kinetic energy settings (centered around 620 eV) imply that we detect the overlapped signals from the 2p$_{1/2}$ and 2p$_{3/2}$ levels of Ar at 250.6 and 248.4 eV. 

\par The FEL delivered photons with a repetition rate of 100 Hz and circular polarization. The averaged time profile in the standard pulse mode tends to a square-shaped pulse and therefore the 25 fs RMS, are equivalent to $\sim\,$87 fs FWHM$\;\approx\;$full pulse length. For short pulses (such as in the two-color mode), the averaged time profile tends to a gaussian pulse \cite{prat2023xray}. % and therefore short pulse durations around 3.2 and 6.4 fs RMS correspond to 7.5 and 15 fs FWHM respectively \cite{prat2023xray}. 
The pulse energy was measured with a gas monitor \cite{sorokin2019gm} and the beamline transmission is estimated to be 73\%. The FEL bandwidth is on average 7.8 eV, estimated from the measured spectra in the photon spectrometer without gas in the gas cell. For the intensity scan experiments, the pulse energy is varied with a gas attenuator installed upstream the beamline and filled with N$_2$ \cite{pradervand2023GATT}. The focus size of the X-ray beam estimated from the induced fluorescence of a Ce-YAG screen at the focus position is approximately 2.2x2.2 $\mu$m$^2$ (sigma), limited by the spatial resolution of the imaging system. 

\subsection{Two-color operation} \label{sec:OM2colors}

\par To generate two-color X-ray pulses, we operate in split-undulator, “fresh-slice” configuration \cite{lutman2016fresh,reiche2016two}, which enables an independent control of both colors and time separations ranging from negative time delays up to 500 fs. The temporal separation of the two pulses is varied by adjusting the current in a magnetic chicane located in between the two undulator sections. In order to individually illuminate different hole positions of the mask, the generation of the two-color pulses  \cite{prat2022widely} is spatially (transversally) offset by 100 $\mu$m in the corresponding undulators sections. This specific operation mode has the additional advantage of avoiding cross-talk between the beams due to multi-stage amplification \cite{wang2024millijoule} when operating with spectrally and temporally close FEL pulses, which can lead to artifacts and beams pointing changes at time zero. The tilt of both colors is adjusted to approximate the focus in the sample position while keeping the beams partially separated in the mask. The pulse duration in the two-color mode is expected to remain between few and 10 fs RMS (23.55 fs FWHM).

\subsection{Model summary} \label{sec:OMmodel}

The model accounts for propagation effects and spectral changes of the three incoming SASE beams and for the FWM signal beam, scattered into the phase-matched direction. The corresponding photon fluxes are calculated using a one-dimensional rate equation model \cite{rohringer2009atomic} coupled with a system of rate equations describing the species kinetics and solved by the Runge-Kutta method. 

\par The model considers first and third order polarization terms in the frequency domain \cite{williams1997two} that relate to absorption (XA), leading to core-excitation and direct valence or core ionization, stimulated (XSE) emission as linear processes and stimulated Raman scattering (SXRS) and XFWM signals, proportional to the imaginary part and absolute squared value of the third order polarization, respectively. The considered XFWM processes are: two-color resonant (Raman-type) processes (TC-XFWM), elastic -degenerated- resonant processes (DXFWM) and non-resonant transient gratings (XTG). Spontaneous emission and Auger-Meitner decay from core-hole states are taken into account. The model considers 7 species: Ne, Ne$_{CH}$, Ne$_{VH}$, Ne$^{+}_{2p}$, Ne$^{+}_{2s}$, Ne$^{+}_{CH}$ and Ne$^{2+}$ (where CH = core-excited, VH = valence-excited, $2p$ and $2s$ indicate the ion state).

\newpage 

\begin{appendix}

\section{Supplementary information}\label{secA1}

\normalfont

\subsection{Model description}  \label{sec:SIsignalGeneration}

Propagation - in absorbing media - can be accounted for by solving the wave equation in its linearized form, assuming the slow varying envelop approximation (SVEA) $\abs {\frac{\partial E}{\partial z}} \ll \abs{ k E}$,$\abs {\frac{\partial^2 E}{\partial z^2}} \ll \abs{ k\frac{\partial E}{\partial z}}$ and thus neglecting quadratic terms. Linear and nonlinear polarization are considered as source terms \cite{guo2013generalized,boyd2008nonlinear}: 

 \begin{equation}
 \left( \frac{\partial}{\partial z} -  \frac{1}{c} \frac{\partial}{\partial t} \right) E_{X}(z,t,\omega) =  \frac{i 2 \pi \omega }{c} \left( P_{}^{(1)}(z,t,\omega) + P_{}^{(3)}(z,t,\omega) \right)   
 \end{equation}
\vspace{5pt}

The induced atomic polarization of n'th order is commonly described via the density matrix formalism calculating the expectation value of the dipole operator $P^{(n)}(t) = 2N Tr\{ \hat{\mu} \hat{\rho}^{(n)} \}$. As typical core-hole lifetimes are significantly longer than the oscillation period of the X-ray field $\omega_X $, the Auger-Meitner decay may then be considered as a Markov chain of instantaneous independent interruptions of the atomic density matrix element and for contemporaneous pulse interaction the XFWM cross-sections can be well described using frequency domain $P^{(n)}(\omega) = \int P^{(n)}(t) exp(-i \omega t)dt$ expressions, with $ \Gamma \approx \tau^{-1}_{ch}$ as linewidth parameter \cite{Prior1986impact}. Linear X-ray photon absorption (XA) and stimulated X-ray emission (XSE) are described by first order polarization terms as:

\begin{equation}
P^{(1)}(\omega) \propto Tr \{ \hat{\mu}_{T} \hat{\rho}^{(1)} \} = \frac{i N }{\hbar} \frac{ \mu^2 E_{\omega}} {\tilde{\Delta}} .     
\end{equation}
\vspace{1pt}

With:
\begin{equation}
\tilde{\Delta} \equiv \tilde{\omega}_{eg} - \omega_{1}  
\end{equation}

as the detuning of the incoming radiation $\omega_1$ from the excited state resonance given by the ground ($\ket{g}$) and core-excited ($\ket{e}$) states. The tilde denotes complex frequencies including relaxation terms described by the lifetime of the core-excites states $\Gamma^{-1}$:

\begin{equation}
\tilde{\omega} =\omega - i \Gamma/2  
\end{equation}

In this notation, the linear X-ray absorption cross-section thus corresponds to $\frac{4 \pi \omega}{\varepsilon_0 c} \Im\{P_\omega^{(1)*}E_\omega \}/E_\omega^2  = \sigma^{XA}_{}(\omega) = 2 \lambda f^{''} [m^{2}]$, with $f^{''}$ being the imaginary part of the scattering form factor. 

The third order nonlinearity is the first non-vanishing term in isotropic media and generally relates to two-photon absorption (XTPA), stimulated Raman (XSR), coherent four-wave-mixing (XFWM) or transient grating (XTG) processes. Considering a three level system with states $\ket{g},\ket{e}$ and $\ket{f}$ (a valence-excited state) and transition dipole moments given by $\mu_{eg}=\bra{e}\mu \ket{g}$  and $\mu_{ef}=\bra{e}\mu \ket{f} $, the induced XFWM polarization for three interacting pulses can be described with terms similar to:

\begin{equation}
P^{(3)} (-\omega_p;\omega_1,-\omega_2,\omega_3) \propto \frac{-i N_g}{\hbar^{3}} \frac{ \bra{e}\mu \ket{f}  \bra{f}\mu\ket{e} \bra{e}\mu \ket{g}\bra{g}\mu\ket{e} E_{\omega_1} E^{*}_{\omega_2} E_{\omega_3} \hspace{10pt}}{(\Tilde{\omega}_{eg}-\omega_{1})(\Tilde{\omega}_{ee}-(\omega_{1}-\omega_{2}))(\Tilde{\omega}_{ef}-(\omega_{1}-\omega_{2}+\omega_{3}))} .     
\label{Eq:A1}
\end{equation}
\vspace{5pt}

where $ N_g $ is the population density of the state $\ket{g}$, $\omega_i$ are the instantaneous frequencies of the interacting fields from the SASE pulse $E_{SASE}(\omega_i) =E_{\omega_i} $. Considering two colors only ($\omega_2 = \omega_1 $ and $\omega_3 \rightarrow \omega_2 $), the induced third order polarization for a stimulated emission FWM process at frequency $\omega_2$  may be approximated as \cite{Williams_1997_TC_FWM}:   
\begin{equation}
\begin{split}
P^{(3)}(-\omega_p;\omega_1,-\omega_1,\omega_2) \propto 
\frac{-N_g}{\hbar^{3}} \frac{ \mu^{2}_{eg}(\omega_1)\mu^{2}_{ef}(\omega_2) E_{\omega_1} E^{*}_{\omega_1} E_{\omega_2} \hspace{10pt}}{\tilde{\Delta} \hspace{5pt} \Gamma_e \hspace{5pt} (\tilde{\omega}_{fe} - \omega_2)} + \hspace{40pt}\\
\frac{N_g}{\hbar^{3}} \frac{ \mu^{2}_{eg}(\omega_1)\mu^{2}_{ef}(\omega_2) E_{\omega_1} E^{*}_{\omega_1} E_{\omega_2} \hspace{10pt}}{\tilde{\Delta} \hspace{5pt} (-\tilde{\omega}_{fg} - (\omega_2 +\omega_1)) \hspace{5pt} (\tilde{\omega}_{fe} - \omega_2)} \hspace{30pt}
\end{split}
\label{Eq:A2}
\end{equation}
\vspace{5pt}

Homodyne detected XFWM signals are proportional to the absolute squared value of the induced third order polarization $S_{\text{\tiny XFWM}}(t) \propto  \abs{ P^{(3} }^{2} $, while the stimulated Raman X-ray scattering (SXRS) is defined by the imaginary part. Summarizing the processes considered in the model that contribute directly or indirectly to the XFWM signals:

\begin{equation}
\begin{split}
I_{\text{\tiny XA}}(\omega) \propto  \Im{{E_{\omega}^*P_{\omega}^{(1)} }}   \hspace{55 pt}\\
I_{\text{\tiny XSE}}(\omega,r) \propto  \Im{{E_{\omega}^*P_{\omega}^{(1)} e^{i(\Vec{k}_{1,2,3}-\Vec{k}_{S})\Vec{r}} }} \\
I_{\text{\tiny SXRS}} (\omega,r)\propto  \Im{{E_{\omega}^*P_{\omega}^{(3)} e^{i (\Vec{k}_{1,2,3}-\Vec{k}_{S})\Vec{r}}}} \\
%\vspace{-10pt} \\
I_{\text{\tiny XFWM}} (\omega,r)\propto  \abs{P_{\omega}^{(3)}e^{i(\Vec{k}_{1}-\Vec{k}_{2}+\Vec{k}_{3}-\Vec{k}_{S})\Vec{r}}  }^{2} 
\end{split}
\label{eq:SIprocesses}
\end{equation}

With Eqs.$\,$\ref{Eq:A2} and \ref{eq:SIprocesses}, we can calculate the amplitude of resonant Raman-type FWM processes at which both transitions are resonant with the core-excited state, \textit{i.e.}, two-color resonant XFWM (TC-XFWM) and stimulated Raman X-ray scattering (SXRS). When $\omega_{1}\,=\,\omega_{2}$ and $\omega_{eg}\,=\,\omega_{ef}$  Eq.$\,$\ref{Eq:A2} describes an elastic -degenerated- resonant XFWM, denoted as DXFWM. From Eq.$\,$\ref{Eq:A2} we define a common nonlinear gain-function for the resonant XFWM processes:

\begin{equation}
g_{\text{\tiny XFWM}}(\omega) =  \frac{P_{\omega}^{(3)}}{N_g E_{\omega} E^{}_{\omega} E_{\omega}} \propto \chi^{(3)}   
\label{Eq:g_fwm_g1}
\end{equation}
\vspace{1 pt}

\par Interference in the non-collinear `BoxCARS' configuration causes a spatial modulation of the ion density in the focal region, an X-ray-induced transient grating (XTG). Bragg-diffracted photons from the incoming beams that are scattered from the XTG propagate together with resonant XFWM contributions into the same phase matched direction. The XTG contribution is not resonant in the `probe' pulse interaction and consists of multiple terms. The corresponding gain factor can be written as: 

\vspace{1 pt}
\begin{equation}
    g_{\text{\tiny XTG}}(\omega) \approx  \frac{1}{\hbar^{3}}  \frac{ \mu^{2}_{eg}(\omega)  \mu^{2}_{ce'} }{\Delta \Gamma_{e} \Gamma_{c}} \bigg\rvert_{\text{\tiny 1s$^{\text{\tiny -1}}$}}+ \frac{1}{\hbar^{3}} \frac{\mu^{4}_{cg}\hspace{10pt}}{ \Gamma_{c}^{3}} \bigg\rvert_{\stackrel{\text{\tiny 2s$^{\text{\tiny -1}}$}}{\text{\tiny 2p$^{\text{\tiny -1}}$}}}     
\label{Eq:g_TG_g2}
\end{equation}
\vspace{1 pt}

The first term in Eq.$\,$\ref{Eq:g_TG_g2} accounts for ions produced by Auger-Meitner decay after resonant core excitation whereas the second term relates to valence ionized Neon, where $\mu_{ic}$ accounts for nonresonant transitions to the ion continuum. $\Gamma_{c}$ may be approximated via the inverse interaction time of the SASE pulse. The nonresonant XTG is a so-called `secondary' grating, at which the induced change of refractive index is given by incoherent relaxation via A-M decay or by direct ionization from valence orbitals. In principle, we could consider a first order core-excited population grating, prior to ionization. However, the nonresonant interaction on the probe make this process significantly weaker than the fully resonant XFWM case (Eq.$\,$\ref{Eq:A2} with $\omega_1\,=\,\omega_2$), and  therefore it has been neglected in the performed calculations. Transient gratings formed in time scales longer than the pulse duration, \textit{e.g.}, induced by heat, pressure, etc., do not contribute relevantly to the signal in our experimental conditions, and are also neglected. Note that for the multiple ions $(Ne^+, Ne^{2+}, ... )$ present in the focal region, the total polarization is approximated as a sum of contributions from the individual species $(S)$ and transitions ($T_{s}$):

\begin{equation}
P^{(3)}(\omega,t) = \sum_{S,T_{s}} N_{s}(t) \tilde{P}_{T_{s}}^{(3)}(\omega)  
\label{Eq:rho_ion}
\end{equation}

Scattered photons from concomitant ion-XTGs can superimpose with the resonantly modified third order sample polarization. The total XFWM signal in the phase matched direction is thus a coherent sum of two terms representing a resonant (RXFWM) and a `nonresonant' (XTG) contribution.

\begin{equation}
\Vec{P}^{(3)}_{\text{\tiny XFWM}}(\omega,\Vec{r})  \approx  \left[ P_{\text{\tiny RXFWM}}^{(3)}(\omega) + e^{i\phi}P_{\text{\tiny XTG}}^{(3)}(\omega) \right] e^{-i(\Vec{k}_{1}-\Vec{k}_{2}+\Vec{k}_{3})\Vec{r}} 
\label{Eq:Polar_sub}
\end{equation}  
\vspace{1 pt}

with $\phi=\omega \tau + \Delta \phi_{0}$ accounting for optical interference of coherent photons with emission delay $\tau$. Strong XTGs can thus enhance resonant XFWM signals similar to heterodyning with a local oscillator field and, as such, mixed terms representing a heterodyne signal have been included in the calculations.

\par Linear photon absorption (XA), stimulated X-ray emission (XSE) and stimulated Raman (SXRS) processes modify the spectral density of each individual pulse in forward beam direction ($\vec{k}_{1,2,3}$). In contrast, the XFWM signals are scattered into the phase-matched direction given by $\Vec{k}_{1}-\Vec{k}_{2}+\Vec{k}_{3}$ in the Boxcars geometry. The model therefore accounts for the propagation of the three incoming SASE beams, considered identical, and for the signal beam, which builds up on the photons scattered from the SASE beams via FWM processes. The propagation of all beams is modelled as photon flux with a one-dimensional rate equation model \cite{rohringer2009atomic}. Beam dispersion effects are neglected and all frequencies propagate with the speed of light:

\begin{equation}
\begin{split}
\frac{\partial I_{\text{\tiny SASE,XFWM}}(\omega,t,z)}{\partial t} =   \sum_{\text{\tiny S,$T_{S}$}}
\pm N_{\text{\tiny S}}(t) \sigma^{\text{\tiny $T_S$}}_{\text{\tiny XA,XSE}}(\omega)cI_{\text{\tiny SASE,XFWM}}(\omega,t,z)  \hspace{170pt} \\
\pm N_{\text{\tiny S}}(t) 
\begin{bmatrix}
g^{\text{\tiny $T_S$}}_{\text{\tiny XNL}}(\omega)I_{\text{\tiny SASE,XFWM}}(\omega,t,z)
\end{bmatrix}
cI_{\text{\tiny SASE,XFWM}}(\omega,t,z)  \hspace{120pt}\\ 
\pm N_{\text{\tiny S}}(t)
\begin{bmatrix}
 N_{\text{\tiny S}}(t) (g^{\text{\tiny $T_S$}}_{\text{\tiny XNL}}(\omega))^2 c t I_{\text{\tiny SASE}}^{2}(\omega,t,z) + \\
 \\
 N_{\text{\tiny S}}(t) (g^{\text{\tiny $T_S$}}_{\text{\tiny XTG}}(\omega))^2 c t I_{\text{\tiny SASE}}^{2}(\omega,t,z)  + \\  
 \\
\hspace{2pt} N_{\text{\tiny S}}(t) g^{\text{\tiny $T_S$}}_{\text{\tiny XNL}}(t,\omega)g^{\text{\tiny $T_S$}}_{\text{\tiny XTG}}(\omega) c t I_{\text{\tiny SASE}}^{2}(\omega,t,z) \\
\end{bmatrix}
 c I_{\text{\tiny SASE}}(\omega,t,z) \hspace{90pt} \\
 + \frac{\phi}{4\pi}N_{\text{\tiny S}}(t)A^{\text{\tiny $T_S$}}c \mp c\frac{\partial I_{\text{\tiny SASE,XFWM}}(\omega,t,z)}{\partial z} \hspace{180pt}
\label{eq:SIpropagation}
\end{split}
\end{equation}

Where the first term accounts for absorption (leading to valence or core ionization and core-excitation) and stimulated emission; the second term accounts for stimulated Raman scattering (SXRS) in beam propagation and the third term represents the four wave mixing processes; the forth term represents the spontaneous emission from core excited states, with $\phi$ as the emission angle and $A_{\text{\tiny T}}$ the corresponding Einstein coefficient for the spontaneous emission. The signs indicate either depletion or addition of photons at the corresponding frequencies in the propagating beams. Note that the terms in the square brackets describe a cross-section for the nonlinear scattered X-ray intensity. Corrections of the cross sections according to the statistical weights of the involved energy levels, no explicitly represented in Eq.$\,$\ref{eq:SIpropagation} were taken into account. 

The beam propagation scheme is coupled to the atomic level kinetics described by a system of rate equations. Indeed, the populations of different species vary within the duration of the applied SASE pulses and, as such, varies their contribution to the detected signals. To describe the abundance of different ion species the time dependent population changes within the SASE pulse duration are calculated according to:

\begin{equation}
\begin{split}
\frac{\partial N^g_{\text{\tiny S}}}{\partial t} = \sum_\text{\tiny $ST_S$} - N^g_{\text{\tiny S}}(\tilde{t}) \sigma^{\text{\tiny XA}}_{\text{\tiny $T_S$}} I_{\text{\tiny SASE,XFWM}}(\omega,\tilde{t},z) + N^e_{\text{\tiny S}}(\tilde{t})\sigma^{\text{\tiny XSE}}_{\text{\tiny $T_S$}} I_{\text{\tiny SASE,XFWM}}(\omega,\tilde{t},z)  \\
+N^e_{\text{\tiny S}}(\tilde{t}) A_{\text{\tiny T}}^{\text{\tiny S}} + N^e_{e}(\tilde{t})\Gamma_{\text{\tiny AM}} 
 \pm N_{\text{\tiny S}}(\tilde{t}) \left[
 g^{\text{\tiny $T_S$}}_{\text{\tiny XFWM}}(\omega)I_{\text{\tiny SASE,XFWM}}(\omega,t,z)
 \right]I_{\text{\tiny SASE,XFWM}}(\omega,t,z) 
\end{split}
\label{eq:SIrateDensities}
\end{equation}

\vspace{10pt}

where $\tilde{t} = t-cz$. The model considers seven population levels $S$ related to different species: (1) Ne, (2) Ne$_{CH}$, (3) Ne$_{VH}$,(4) Ne$^{+}_{2p}$, (5) Ne$^{+}_{2s}$, (6) Ne$^{+}_{CH}$ and (7) Ne$^{2+}$ (where CH = core-excited, VH = valence-excited, $2p$ and $2s$ indicate the ion state). Higher ionization states Ne${^\text{\tiny >2+}}$ are neglected. The differential equations represented in Eq.$\,$\ref{eq:SIrateDensities} account in general for linear absorption leading to core-excitation and direct valence or core ionization, stimulated and spontaneous emission and Auger-Meitner decay from core-hole states, but also stimulated Raman scattering, which changes the population of two specific levels, Ne and Ne$_{VH}$. The term $g^{\text{\tiny $T_S$}}_{\text{\tiny XFWM}}$ therefore only applies for the TC-XFWM and SXRS cases. Obviously not all terms contribute to all population levels, and Eq.$\,$\ref{eq:SIrateDensities} is adapted accordingly.

\par The system of rate equations describing the species kinetics, coupled to the beams propagation and amplification are solved by the Runge-Kutta method. The stochastic SASE pulses from the X-FEL laser source are simulated according to Ref.$\,$\cite{pfeifer2010partial}. 

\newpage 

\subsection{Supplementary model results} \label{sec:SImodelresults} 

\begin{figure}[h]%
	\centering
	   % \fbox{
  	\includegraphics[width=0.49\textwidth]{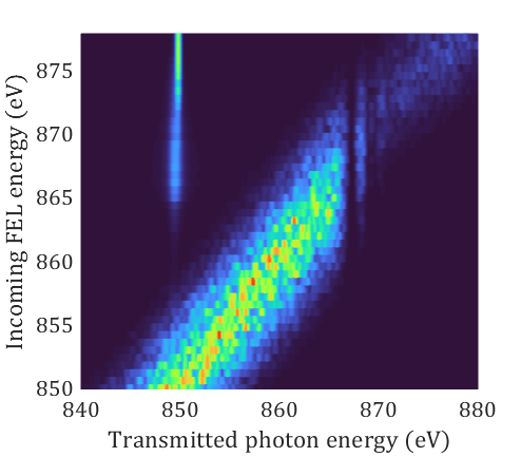}
    \includegraphics[width=0.49\textwidth]{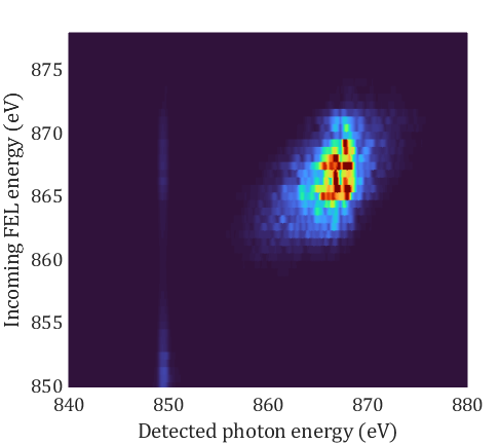}
    % \fbox{
  	\includegraphics[width=0.49\textwidth, trim=0cm 0cm 3cm 0cm,clip]{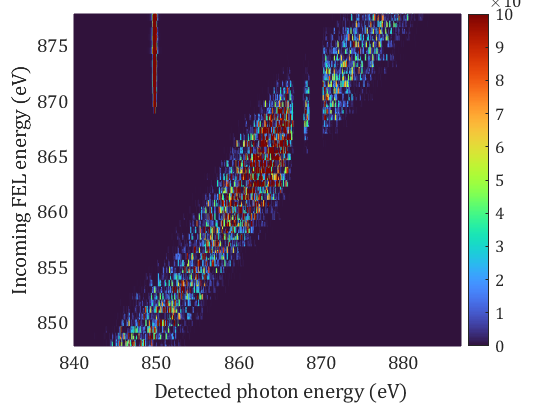}
    % }
	
	\caption{Top left: Simulated spectral map of a single propagating beam for the case of 800 mbar of neon, with 25 fs RMS pulse duration (experimental data and simulation of the XFWM signal in these conditions are shown in figure \ref{fig:Spectra&Fits}$\,$(a) and (b), respectively; Top Right: Simulation of the FWM signal for the 200 mbar, 25 fs RMS case (experimental data shown in figure \ref{fig:Spectra&Fits}$\,$(c)); Bottom: Model of the FWM spectral map obtained imposing a strong transient grating (XTG) contribution.}
	\label{Fig:SImodelPtau}
\end{figure}

The results of the model with a strong grating show a smooth increase of the $\omega_{in}$=$\omega_{out}$ signal as a function of the photon energy, following the ionization degree. On the contrary, the measured signals are characterized by a well defined signal around the neon pre-edge resonances, not reproduced by the model when a strong grating is considered. In addition, a dominant XTG does not show the resonant effect on Ne$^{+}$ below edge ($\omega_{in}$=$\omega_{out}$ = 848.5 eV) (see \ref{sec:SImodelresults}) since valence ionization is constant in this energy range, and no amplified Ne$^{+}$ emission below the ionization threshold is reached. 

\newpage

\subsection{Beam propagation and mask transmission} \label{sec:SImaskTransmission}
 The variation of the intensity distribution along the beam propagation is calculated with direct summation of the Fresnel-Kirchoff diffraction integral converted into a convolution form and computed using a Fast Fourier transform algorithm. The propagation around the focus is implemented using the finite difference method \cite{lightpipes}.
 
\begin{figure}[h]%
	\centering
	   % \fbox{
	\includegraphics[width=1\textwidth]{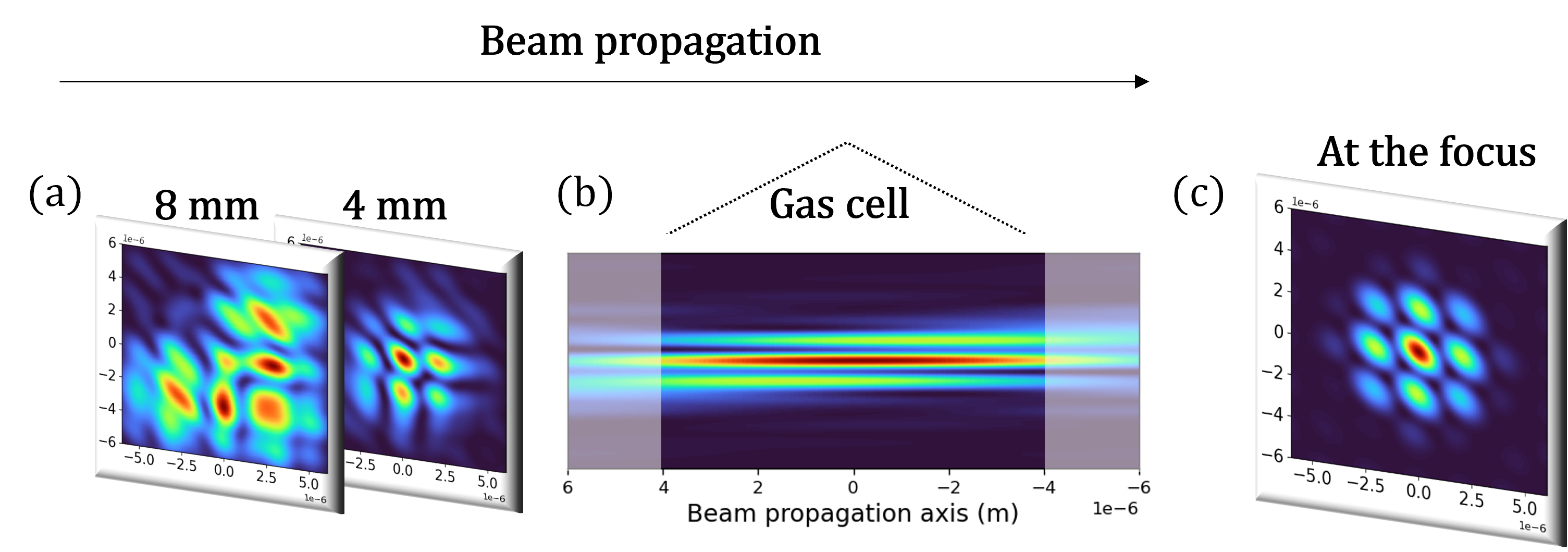}
 % }
	\caption{(a) Calculation of the beam distribution at two positions before entering in the gas cell; (b) Propagation around the focus and intensity distribution at the focus (c).}
	\label{Fig:SIBeamProp}
\end{figure}

 The beam size on the KB mirrors of the Maloja end-station is estimated to be on the order of 3.3$\times$3.2 mm FWHM. The mirrors are 4.5 m away from the focus. The transmission of the spatial mask, placed 1.5 m away from the mirrors, is then calculated assuming an incoming Gaussian pulse to be in the order of 13.7 \%. The intensity distribution close to the entrance of the gas cell, along the gas cell and at the focus is shown in figure \ref{Fig:SIBeamProp}.

\newpage
\subsection{Spatial characterization of the signal} \label{sec:SIyag}

\par
\begin{figure}[h]%
	\centering
	%    \fbox{
	\includegraphics[width=0.32\textwidth]{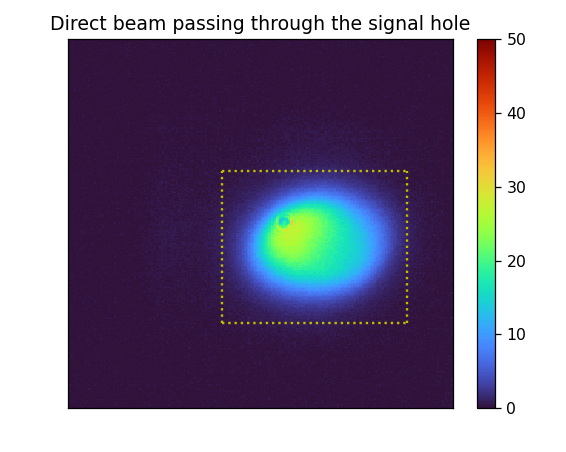}
	\includegraphics[width=0.32\textwidth]{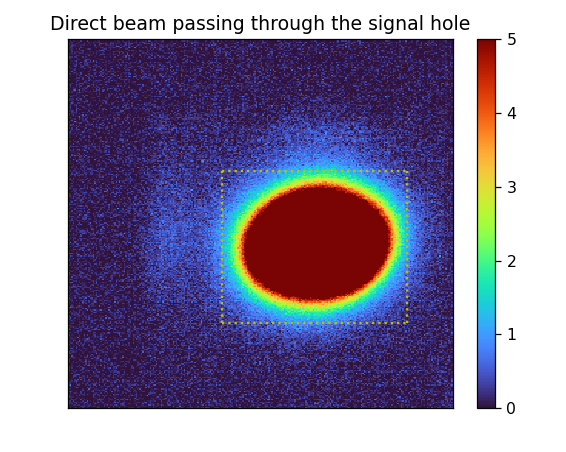}\\
	\includegraphics[width=0.32\textwidth]{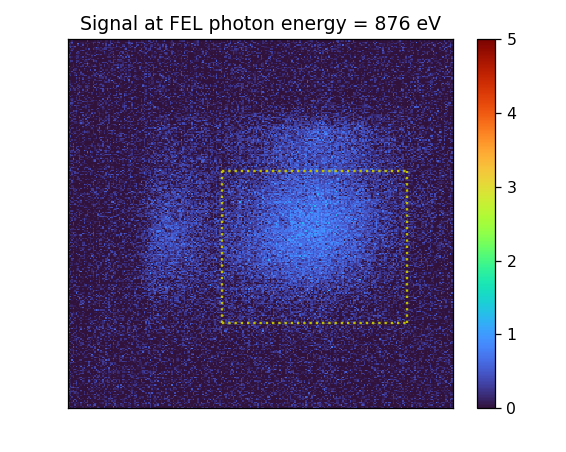}
 	\includegraphics[width=0.32\textwidth]{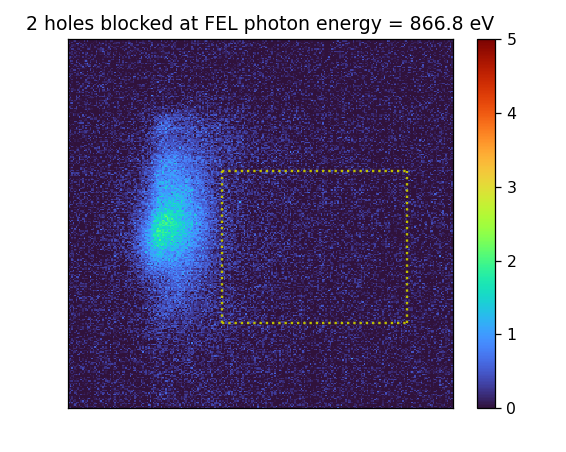}
	%    }
	\caption{Top: Incoming beam opened in the signal direction (4 holes opened in the mask) measured in the YAG with different scaling (to the right scaled similarly as for the signal beams). Bottom signal detected with 3 holes opened in the spatial mask (left) or 2 holes opened (right). All measurements were done with the same gas pressure of 800 mbar. Each image is the average of 2000 FEL shots. Note the different photon energies for the two bottom figures.}
	\label{Fig:SIyagStatic}
\end{figure}

\begin{figure}[h]
	\centering    
	%    \fbox{
	\includegraphics[width=0.32\textwidth]{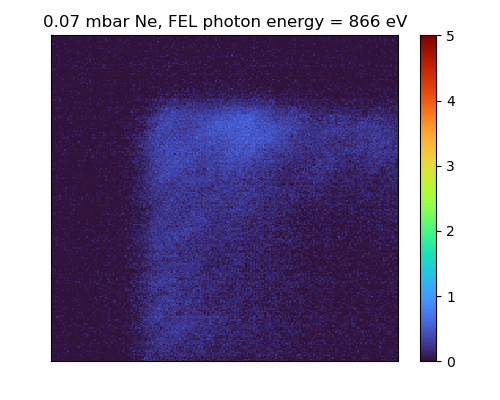}
	\includegraphics[width=0.32\textwidth]{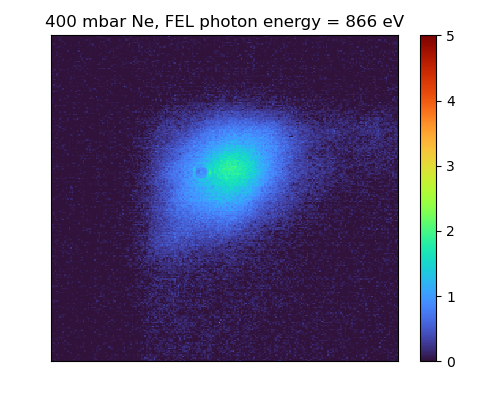}
	\includegraphics[width=0.32\textwidth]{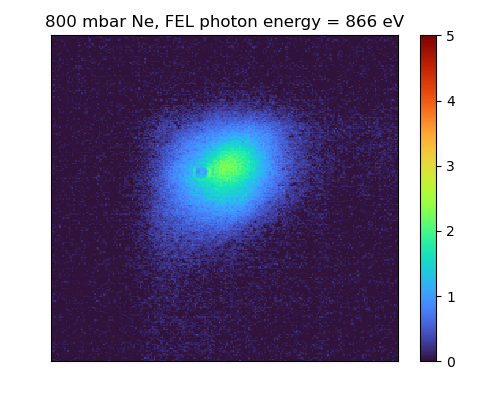}
	%    }
	\caption{Pressure scan imaging the signal beam in the YAG, measured at the same photon energy (866 eV). Each image is the average of 5000 shots. Left: 0.07 mbar; Center: 400 mbar; Right: 800mbar. Note that the imaging system and the positions of the beam blocks that block the three incoming beams were modified between the data shown in figure \ref{Fig:SIyagStatic} and this data set. Note as well the different photon energy compared to the bottom left image in figure \ref{Fig:SIyagStatic}.}
	\label{fig:SIYagPressure}
\end{figure}

 The average of 2000 FEL shots on the YAG screen fluorescence measured with the incoming FEL beam in the direction of the signal (4 holes of the mask opened) and of the signal itself (3 holes opened, hole corresponding to the signal beam blocked) are shown in Fig.$\,$\ref{Fig:SIyagStatic}. The signal shape is the same in both cases. The scatter from the other three incoming beams is visible in the sides of the incoming beam (4 holes open - top-right plot) or the signal (3 holes open - bottom-left plot). A preliminary pressure scan was done imaging the XFWM in the YAG. With a slightly different imaging and signal filtering geometry, the images resulting from the average of 5000 shots are shown in figure \ref{fig:SIYagPressure}. No signal is present when the gas cell is pumped.

\begin{figure}[t]
	\centering
	\includegraphics[width=1\textwidth]{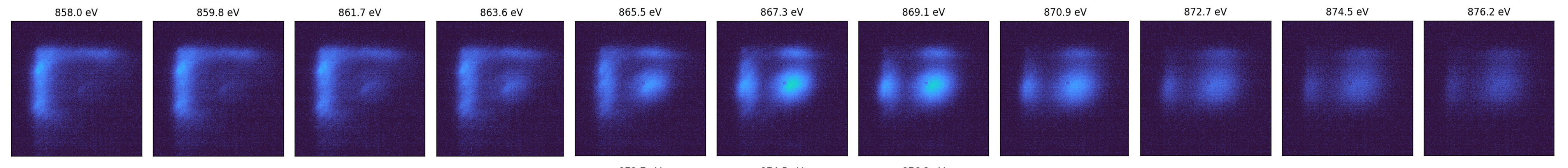}
	\caption{Set of images of the Ce-YAG fluorescence recorded as a function of the incoming photon energy. The measurements were done with the three incoming beams blocked and each image is the average of four independent data sets. The strength of the all X-ray FWM signal shows a clear resonance dependency, with the largest intensity around the pre-edge Ne transitions.}
	\label{fig:SIyagEnergyScan1beam}
\end{figure}

The signal recorded as a function of the photon energy scan is shown in Fig.$\,$\ref{fig:SIyagEnergyScan1beam}.

\newpage

\subsection{Supplementary experimental results on the spectrally resolved XFWM signals} \label{sec:SIspectra}

\subsubsection{FWM verification} \label{sec:SIverification}

Measurements blocking different holes in the spatial mask.

\begin{figure}[h]
	\centering    
	%    \fbox{
	\includegraphics[width=0.45\textwidth]{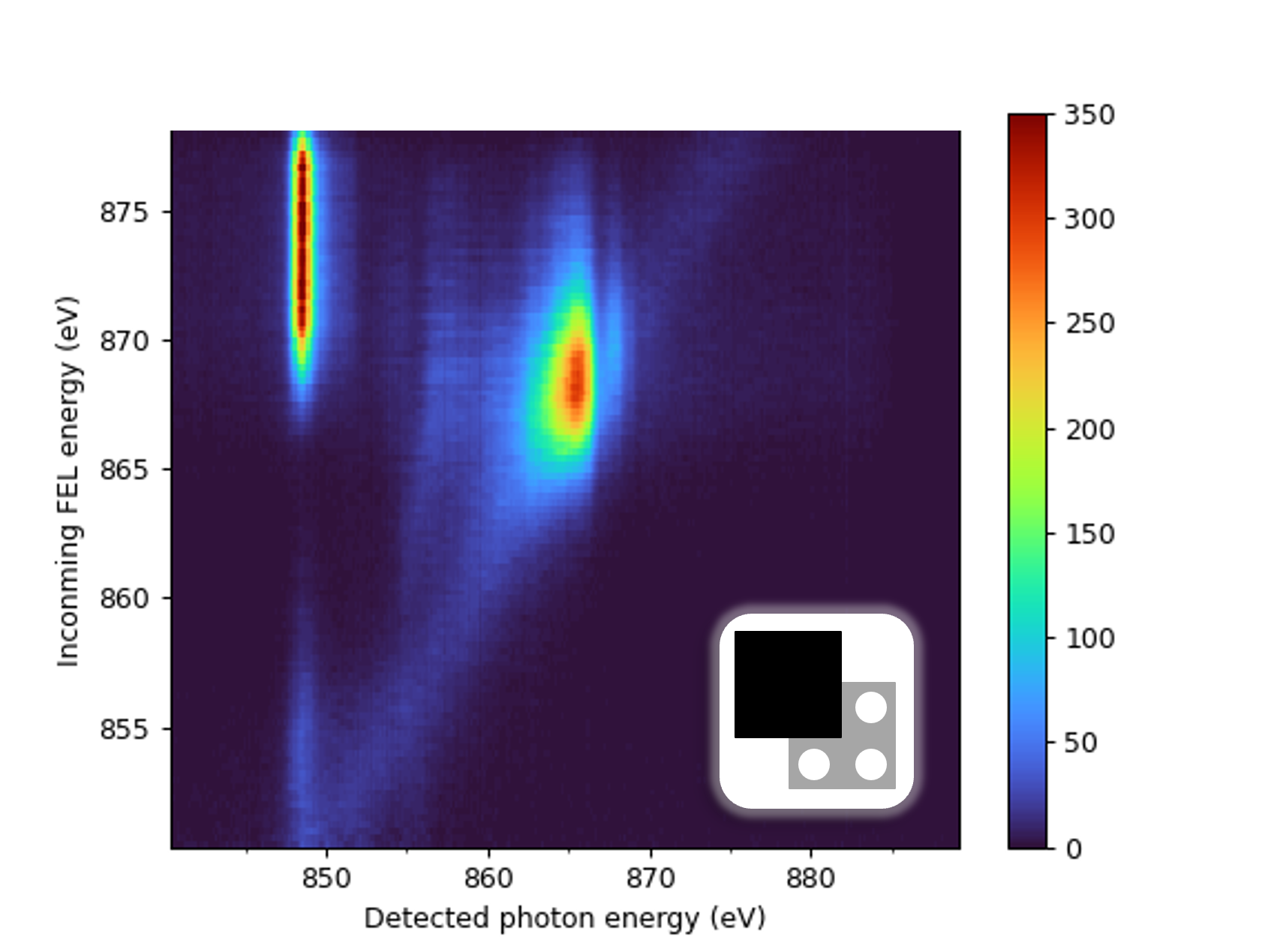} \\
 
	\includegraphics[width=0.45\textwidth]{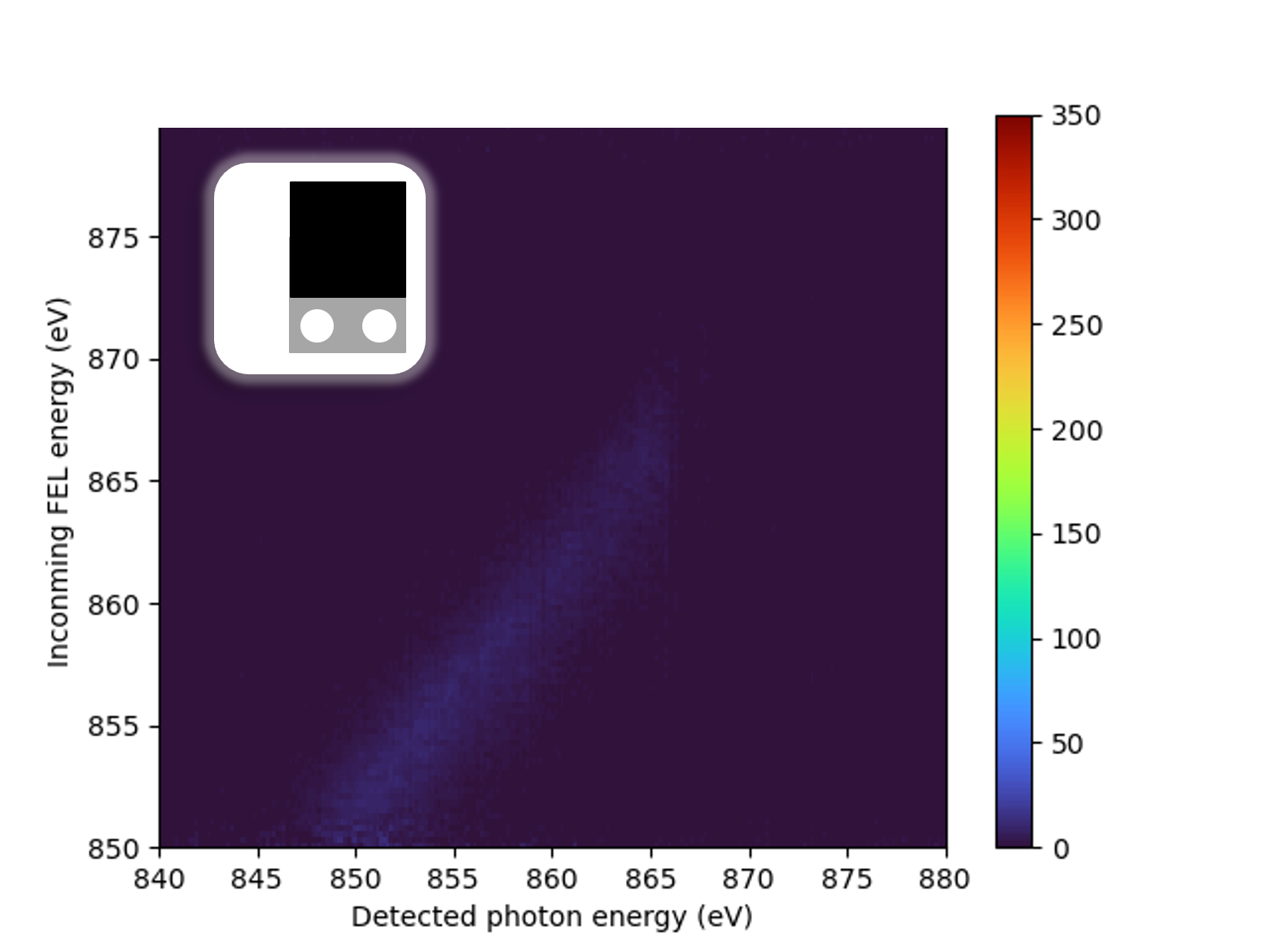}
	\includegraphics[width=0.45\textwidth]{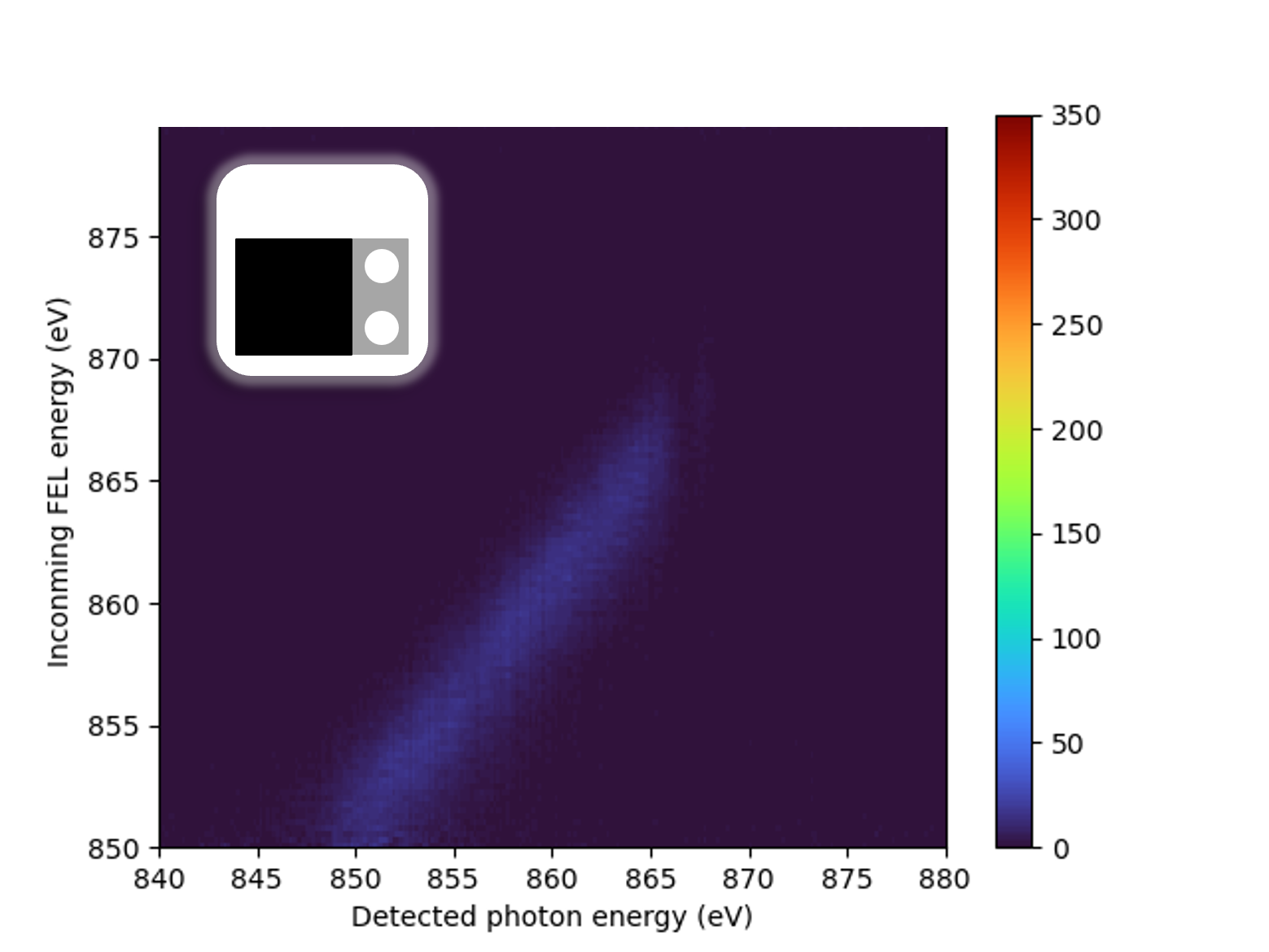}
	%    }
	\caption{Top: standard measurement configuration, \textit{i.e.}, three holes of the spatial mask open, only one incoming beam closed; Bottom: photon energy maps measured closing two of the incoming beams.  An indicative scheme of each mask and beam block configuration is shown as inlet in the three 2D spectral maps. 
    All measurements were done with the same gas pressure of 800 mbar and 25 fs RMS pulse duration.}
	\label{fig:SIholeTests}
\end{figure}

\newpage

\subsubsection{Reference measurement (no gas case)}

\par Measurements done with three incoming beams opened, but with no gas in the gas cell (pumped down to a minimum pressure around 0.1 mbar). These measurement show a larger amount of scattered light than in the measurement shown in Fig.$\,$\ref{fig:SIholeTests} probably because of small changes in the windows surrounding the holes in the gas cell that allow the beams in/out under low pressure conditions. 

\begin{figure}[h]
	\centering
	%    \fbox{
	\includegraphics[width=0.6\textwidth]{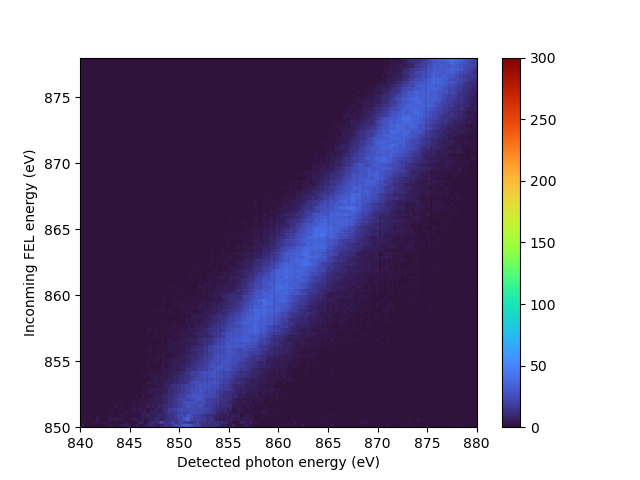}
	%    }
	\caption{Photon energy map measured with three incoming beams and a gas pressure of 0.1 mbar. 
    }
	\label{Fig:noGas}
\end{figure}

\newpage
\subsubsection{Pressure variation} \label{sec:SIpressure}

Comparison of the photon energy scans done for different pressures (200, 400 and 800 mbar) done with the standard pulse duration, 25 fs RMS, with full FEL transmission. Each row is roughly 600 shots.

\begin{figure}[h]%
	\centering
 	   % \fbox{
     \includegraphics[width=0.49\textwidth]{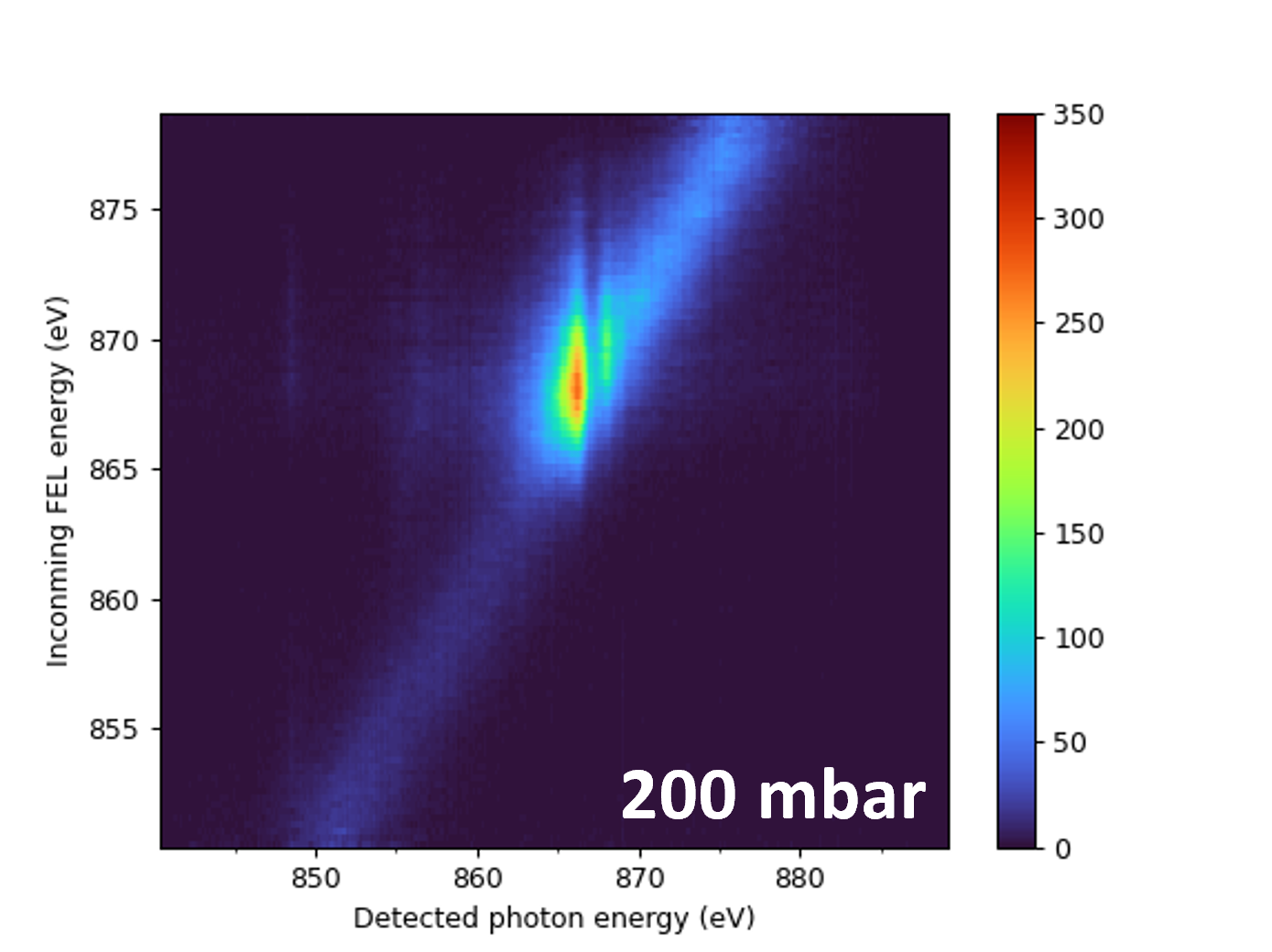}        
     \includegraphics[width=0.49\textwidth]{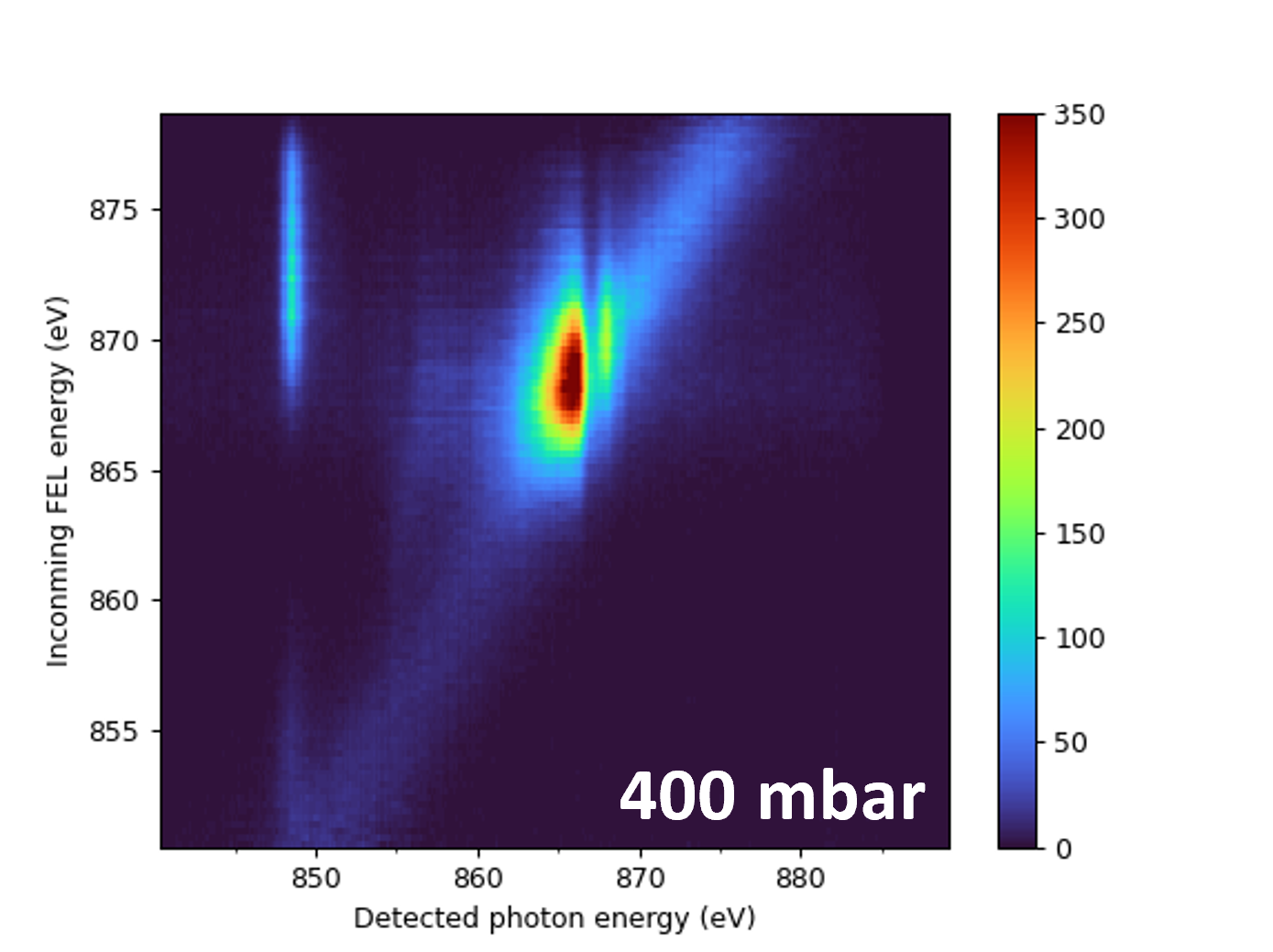}         
     \includegraphics[width=0.49\textwidth]{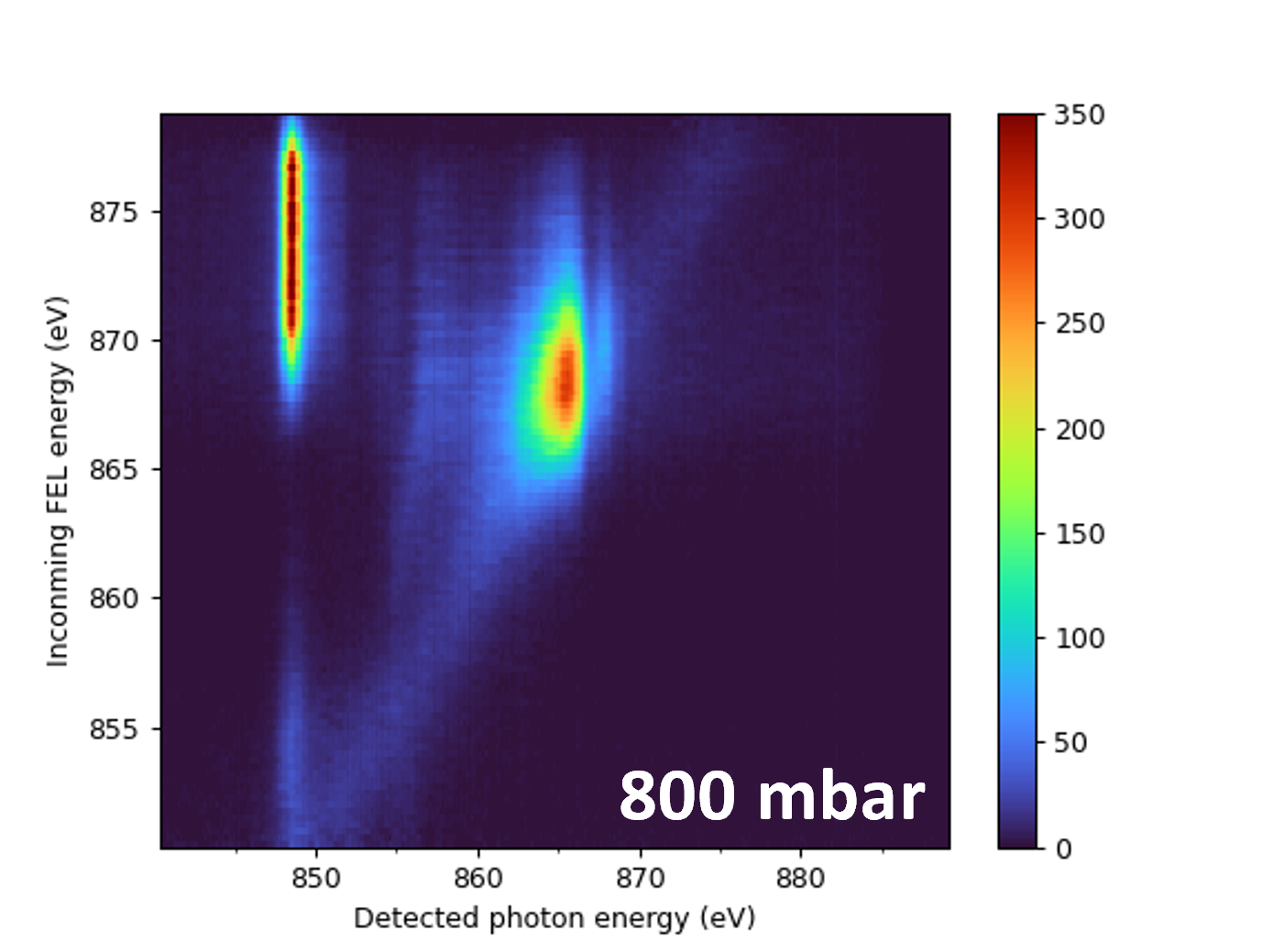}
 	   % }

	\caption{Photon energy scans for 200 (top left), 400 (top right) and 800 mbar (bottom) measured with a pulse duration of 25 fs RMS plotted as spectral maps as a function of the incoming and detected photon energies. 
   }
	\label{Fig:SIPressureVariation}
\end{figure}

\newpage
\subsubsection{Intensity variation} \label{sec:SIintensity}

\par FEL intensity scans were done for two pressures, 800 and 200 mbar, and two photon energies: 866.4 and 876.6 eV (average of the center photon energy obtained from the gaussian fit of all step scans in each data set). The FEL intensity was varied by changing the transmission of the FEL beam with a gas attenuator from 0.1 to 1 (full transmission). 

\begin{itemize}

\item \textbf{800 mbar - Intensity scan at high energy (876.6 eV)}

\begin{figure}[h]%
	\centering
	%    \fbox{
	\includegraphics[width=0.45\textwidth]{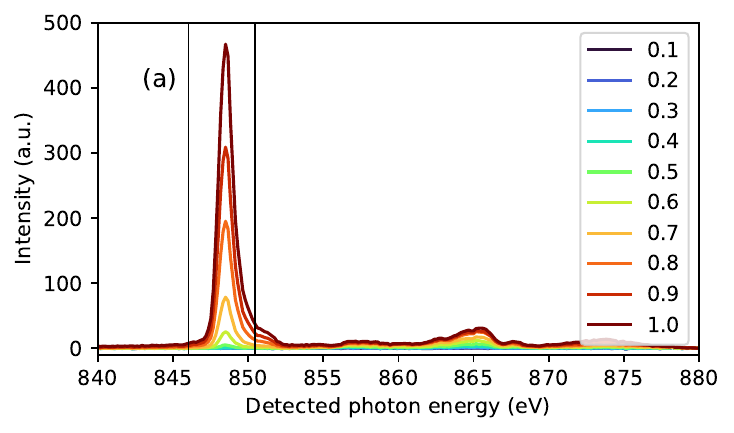}
	%    }
 	\includegraphics[width=0.45\textwidth]{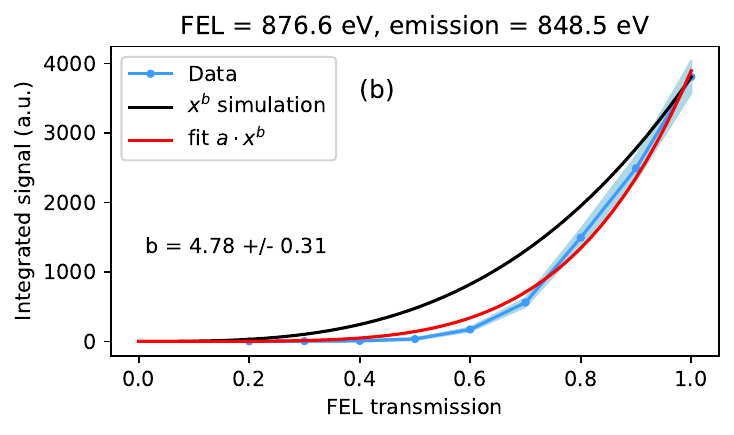} 
	\caption{(a) Spectra measured for different FEL transmissions at 800 mbar, 876.6 eV FEL photon energy and 25 fs RMS pulse duration. The black lines show the ROI taken for the fit; (b) Fit of the integrated signals as a function of the intensity (FEL transmission).}
	\label{Fig:SIFELintensityHighEnergy}
\end{figure}

\item \textbf{800 mbar - Intensity scan at resonance (866.4 eV)}

\begin{figure}[h]%
	\centering
	%    \fbox{
	\includegraphics[width=0.5\textwidth]{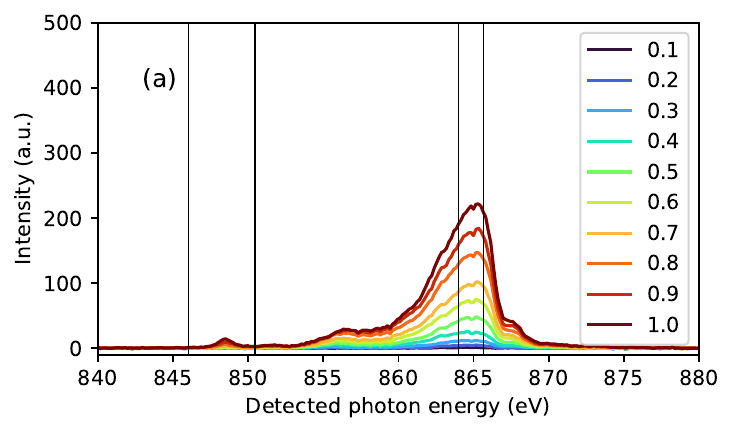} \\
	%    }
 	\includegraphics[width=0.4\textwidth]{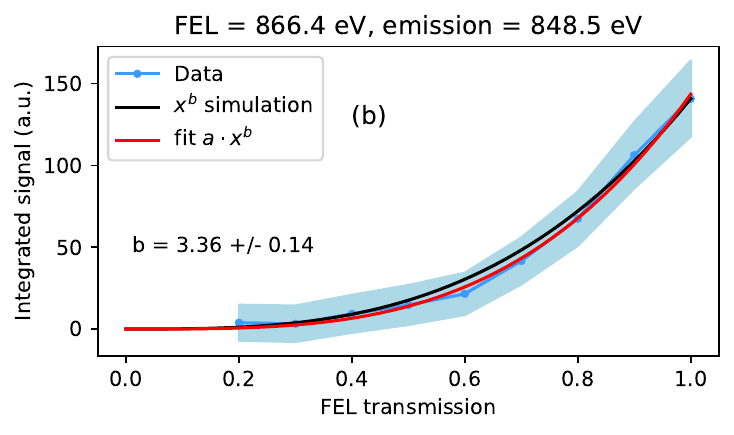} 
   	\includegraphics[width=0.4\textwidth]{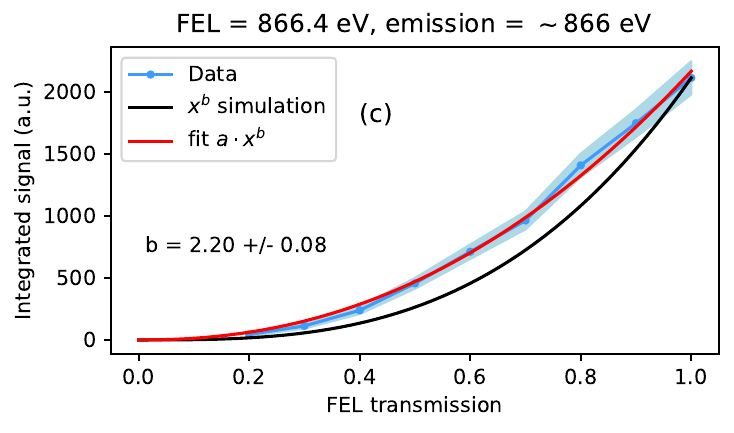} 
    	\caption{(a) Spectra measured for different FEL transmissions at 800 mbar, 866.4 eV FEL photon energy and 25 fs RMS pulse duration. The black lines show the ROIs taken for the two fits shown in panels (b) and (c).}
	\label{Fig:SIFELintensityRes}
\end{figure}

\newpage

\item\textbf{200 mbar - Intensity scan at resonance (866.8 eV)}

\begin{figure}[h]%
	\centering
	%    \fbox{
	\includegraphics[width=0.55\textwidth]{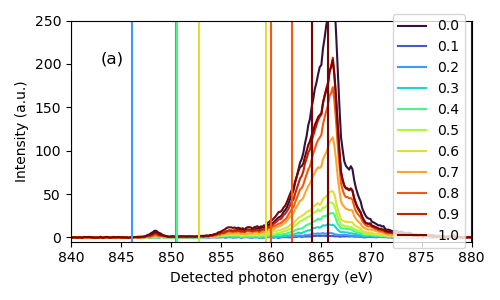}
	%    }
 	\includegraphics[width=0.49\textwidth]{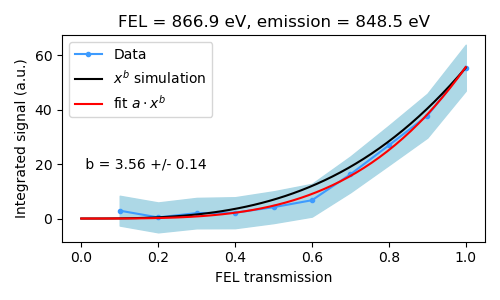} 
   	\includegraphics[width=0.49\textwidth]{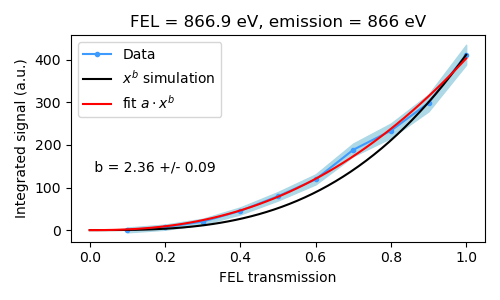} 
    \caption{(a) Spectra measured for different FEL transmissions at 200 mbar, 866.8 eV FEL photon energy and 25 fs RMS pulse duration. The coloured lines show the ROIs taken for the fit (first and last); (b) and (c) Fits of the integrated signals as a function of the intensity (FEL transmission).}
	\label{Fig:SIFELintensityShortPulse}
\end{figure}

\end{itemize}

\newpage

\subsubsection{Low incoming photon energy data} \label{sec:SIlowEnergy}

\par Photon energy scan taken at 800 mbar Ne for two photon energy ranges to explore the ion emission below the Ne K edge. These measurements were taken before the final scattering optimization, explaining the larger scattering when compared with the data taken in similar experimental conditions but with optimized scattering reduction (see \ref{sec:SIpressure}). 

\begin{figure}[h]%
	\centering
	%    \fbox{
	\includegraphics[width=0.65\textwidth]{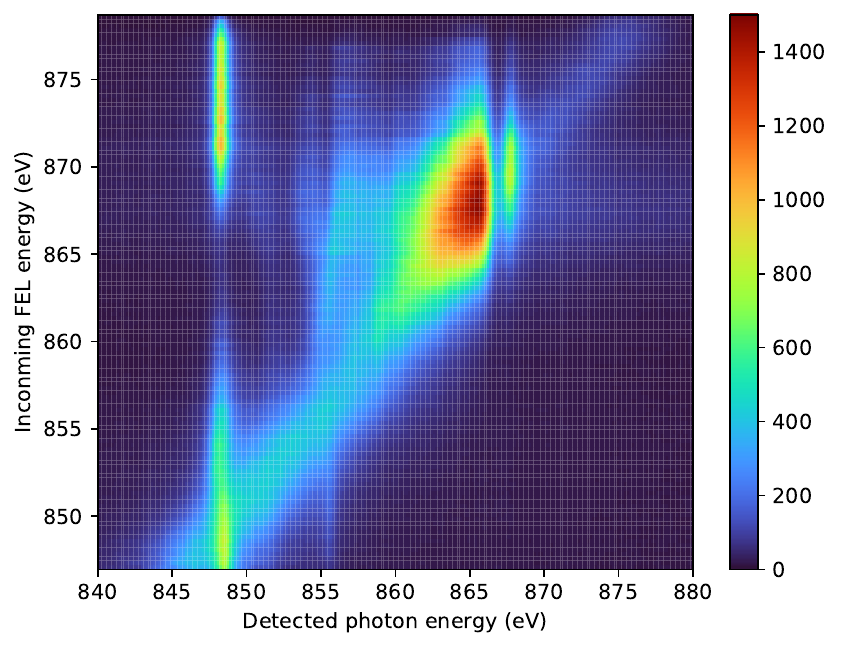}
 	\includegraphics[width=0.95\textwidth]{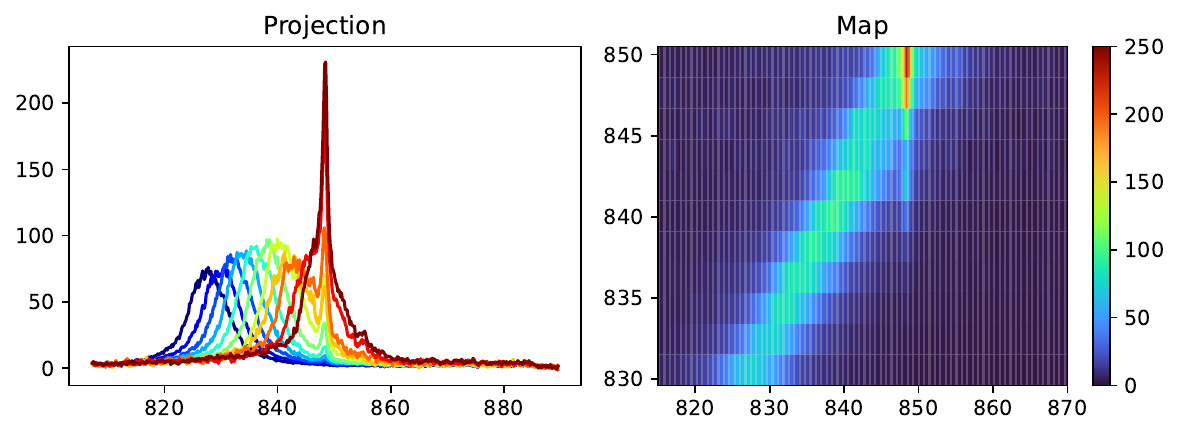}
	%    }
	\caption{Photon energy scans for the usual (top) and lower (bottom) FEL photon energy range taken in the same conditions. No electron spectrometer data is available in the latter case and therefore the data could not be energy-sorted.}
	\label{Fig:SIlowEnergy}
\end{figure}

\end{appendix}

\newpage

\bibliographystyle{newstyle}
\bibliography{All_X_ray_FWM_arxiv}  

\end{document}